\documentclass[12pt]{article}
\usepackage{latexsym,amssymb,amsmath}
\begin{document}
\baselineskip=20pt

\newcommand{\la}{\langle}
\newcommand{\ra}{\rangle}
\newcommand{\psp}{\vspace{0.4cm}}
\newcommand{\pse}{\vspace{0.2cm}}
\newcommand{\ptl}{\partial}
\newcommand{\dlt}{\delta}
\newcommand{\sgm}{\sigma}
\newcommand{\al}{\alpha}
\newcommand{\be}{\beta}
\newcommand{\G}{\Gamma}
\newcommand{\gm}{\gamma}
\newcommand{\vs}{\varsigma}
\newcommand{\Lmd}{\Lambda}
\newcommand{\lmd}{\lambda}
\newcommand{\td}{\tilde}
\newcommand{\vf}{\varphi}
\newcommand{\yt}{Y^{\nu}}
\newcommand{\wt}{\mbox{wt}\:}
\newcommand{\rd}{\mbox{Res}}
\newcommand{\ad}{\mbox{ad}}
\newcommand{\stl}{\stackrel}
\newcommand{\ol}{\overline}
\newcommand{\ul}{\underline}
\newcommand{\es}{\epsilon}
\newcommand{\dmd}{\diamond}
\newcommand{\clt}{\clubsuit}
\newcommand{\vt}{\vartheta}
\newcommand{\ves}{\varepsilon}
\newcommand{\dg}{\dagger}
\newcommand{\tr}{\mbox{Tr}}
\newcommand{\ga}{{\cal G}({\cal A})}
\newcommand{\hga}{\hat{\cal G}({\cal A})}
\newcommand{\Edo}{\mbox{End}\:}
\newcommand{\for}{\mbox{for}}
\newcommand{\kn}{\mbox{ker}}
\newcommand{\Dlt}{\Delta}
\newcommand{\rad}{\mbox{Rad}}
\newcommand{\rta}{\rightarrow}
\newcommand{\mbb}{\mathbb}
\newcommand{\lra}{\Longrightarrow}
\newcommand{\X}{{\cal X}}
\newcommand{\Y}{{\cal Y}}
\newcommand{\Z}{{\cal Z}}
\newcommand{\U}{{\cal U}}
\newcommand{\V}{{\cal V}}
\newcommand{\W}{{\cal W}}

\begin{center}{\Large \bf Asymmetric and Moving-Frame  Approaches }\end{center}
\begin{center}{\Large \bf to Navier-Stokes Equations}
\end{center}
\vspace{0.2cm}

\begin{center}{\large Xiaoping Xu}\end{center}
\begin{center}{Institute of Mathematics, Academy of Mathematics \& System Sciences}\end{center}
\begin{center}{Chinese Academy of Sciences, Beijing 100080, P.R. China}
\footnote{Research supported
 by China NSF 10431040}\end{center}

\vspace{0.6cm}

 \begin{center}{\Large\bf Abstract}\end{center}

\vspace{1cm} {\small In this paper, we introduce a method of
imposing asymmetric conditions on the velocity vector with respect
to independent variables and a method of moving frame for solving
the three dimensional Navier-Stokes equations. Seven families of
non-steady rotating asymmetric solutions with various parameters
are obtained. In particular, one family of solutions blow up at
any point on a moving plane with a line deleted, which may be used
to study turbulence. Using Fourier expansion and two families of
our solutions, one can obtain discontinuous solutions that may be
useful in study of shock waves. Another family of solutions are
partially cylindrical invariant, contain two parameter functions
of $t$ and structurally depend on two arbitrary polynomials, which
may be used to describe incompressible fluid in a nozzle. Most of
our solutions are globally analytic with respect to spacial
variables.}

\section{Introduction}

The most fundamental differential equations in the motion of
incompressible viscous fluid are Navier-Stokes equations:
$$u_t+uu_x+vu_y+wu_z+\frac{1}{\rho}p_x=\nu (u_{xx}+u_{yy}+u_{zz})
 ,\eqno(1.1)$$
$$v_t+uv_x+vv_y+wv_z+\frac{1}{\rho}p_y=\nu (v_{xx}+v_{yy}+v_{zz})
 ,\eqno(1.2)$$
$$w_t+uw_x+vw_y+ww_z+\frac{1}{\rho}p_z=\nu (w_{xx}+w_{yy}+w_{zz})
 ,\eqno(1.3)$$
$$u_x+v_y+w_z=0,\eqno(1.4)$$
 where
$(u,v,w)$ stands for the velocity vector of the fluid, $p$ stands
for the pressure of the fluid, $\rho$ is the density constant and
$\nu$ is the coefficient constant of the kinematic viscosity.

The Lie point symmetries of the two-dimensional special case of
the above equations ($u_z=v_z=w=0$) were obtained by Pukhnachev
[P1] and Buchnev [Ba]. Moreover, certain group-invariant solutions
were found in the works of Pukhnachev [Pv1], Kochin-Kibel'-Roze
[KKR] and Bytev [Bv1], [Bv2]. Futhermore, Gryn [G] obtained
certain exact solution describing flows between porous walls in
the presence of injection and suction at identical rates, and
Polyanin [Pa] used the method of generalized separation of
variables to find certain exact solutions.

Assuming nullity of certain components of the tensor of momentum
flow density, Landau [Ll] found a exact solution of Navier-Stokes
equations (1.1)-(1.4), which describes axially symmetrical jet
discharging from a thin pipe into unbounded space. The Lie point
symmetries of the above three-dimensional equations were obtained
by Buchnev [Ba] and Pukhnachev [Pv2]. Moreover, Kapitanskii [K]
found certain cylindrical invariant solutions of the equations and
Yakimov [Y] obtained exact solutions with a singularity of the
type of a vortex filament situated on a half line. Shen [S1, S2]
rewrote Navier-Stokes equations in terms of complex variables and
found certain exact solutions. Brutyan and Karapivskii [BK] got
exact solutions describing the evolution of a vortex structure in
a generalized shear flow. Furthermore, Leipnik [Lr] obtained exact
solutions by recursive series of diffusive quotients, and
Vyskrebtsov [V] studied self-similar solutions for an axisymmetric
flow of a viscous incompressible flow.

From algebraic point of view, it seems to us that  there are not
enough exact solutions that fully reflect the fundamental natures
of Navier-Stokes equations. In this paper, we introduce a method
of imposing asymmetric conditions on the velocity vector with
respect to independent variables and obtain two families of
non-steady asymmetric solutions with rotation. One of the families
contains two arbitrary parameter functions of $t$ and an arbitrary
number of parameter constants. Using Fourier expansion and this
family of solutions, one can obtain discontinuous solutions that
may be useful in study of shock waves. Another family of solutions
are partially cylindrical invariant, contain two parameter
functions of $t$ and structurally depend on two arbitrary
polynomials, which may be used to describe incompressible fluid in
a nozzle. In order to better reflect the rotating nature of flow,
we also give a method of moving frame and find five families of
non-steady rotating solutions with various parameters. In
particular, one family of solutions blow up at any point on a
moving plane with a line deleted, which may be used to study
turbulence. Another family can be used to obtain discontinuous
rotating solutions. Most of our solutions are globally analytic
with respect to spacial variables. Below we give a more detailed
introduction.

The equations (1.1)-(1.4) are invariant under orthogonal
transformations $\{T_A\mid A\in O(n,\mbb{R})\}$ with
$$T_A\left[\left(\begin{array}{c}x\\ y\\ z\end{array}\right)\right]=
\left(\begin{array}{c}x\\ y\\ z\end{array}\right)A,\qquad
T_A\left[\left(\begin{array}{c}u\\ v\\ w\end{array}\right)\right]=
\left(\begin{array}{c}u\\ v\\
w\end{array}\right)A,\qquad T_A(p)=p.\eqno(1.5)$$ Moreover, they
are invariant under the time translation $t\mapsto t+a$  with
$a\in\mbb{R}$ and the scaling $T_b$ with $0\neq b\in\mbb{R}$:
$$T_b(u)=b^{-1}u(b^2t,bx,by,bz),\qquad T_b(v)=
b^{-1}v(b^2t,bx,by,bz),\eqno(1.6)$$
$$T_b(w)=b^{-1}w(b^2t,bx,by,bz),\qquad T_b(p)=
b^{-2}p(b^2t,bx,by,bz).\eqno(1.7)$$ The most interesting
symmetries of Navier-Stokes equations are the following
time-dependent translations:
$$T_{1\al}(u)=u(t,x+\al,y,z)-\al',\qquad
T_{1\al}(v)=v(t,x+\al,y,z),\eqno(1.8)$$
$$T_{1\al}(w)=w(t,x+\al,y,z),\qquad T_{1\al}(p)=
p(t,x+\al,z)+\rho{\al'}'x\eqno(1.9)$$ and its permutations on
$(u,x),\;(v,y),\;(w,z)$, and
$$T_{2\al}(u)=u,\qquad T_{2\al}(v)=v,\qquad T_{2\al}(w)=w,\qquad
T_{2\al}(p)= p+\al,\eqno(1.10)$$ where $\al$ is an arbitrary
function of $t$. The above transformations transform solutions of
Navier-Stokes equations into their solutions. Our goal in this
paper is to find exact solutions of Navier-Stokes equations modulo
the above symmetries. In other words, the above symmetries will be
used to simplify our ansatzes for exact solutions and related
arguments.

For convenience, we always assume that all the involved partial
derivatives of related functions always exist and we can change
orders of taking partial derivatives. In fluid dynamics,
rotation-free solutions of Navier-Stokes equations, namely,
$$u_y-v_x=0,\qquad v_z-w_y=0,\qquad w_x-u_z=0,\eqno(1.11)$$
are not so interesting. From pure mathematical point of view, a
rotation-free solution is equivalent to a time-dependent harmonic
function $f(t,x,y,x)$ (i.e., $f_{xx}+f_{yy}+f_{zz}=0$), where
$$u=f_x,\qquad v=f_y,\qquad w=f_z.\eqno(1.12)$$
Practically, steady solutions (or time-independent) are not very
important. In general, it is difficult to find exact non-steady
rotating solutions for Navier-Stokes equations (1.1)-(1.4) due to
their nonlinearity.

Using certain finite-dimensional stable range of the nonlinear
term, we found in [X1] a family of exact solutions with seven
parameter functions for the equation of nonstationary transonic
gas flows found by Lin, Reisner and Tsien [LRT], which  blow up on
a moving line. These solutions may reflect partial phenomena of
gust. In [X2], we use various ansatzes with undermined functions
and the technique of moving frame to find basic solutions modulo
the Lie point symmetries with parameter functions for the
classical non-steady boundary layer problems. These two works
motivated us to solve Navier-Stokes equations by algebraic
methods.

Our first idea is to impose suitable asymmetric conditions on the
velocity vector with respect to independent variables. For
instance, assuming
$$u=\gm(t)x+y\phi(t,x^2+y^2),\;\;v=\gm(t)y-x\phi(t,x^2+y^2),\;\;w=
\psi(t,x^2+y^2)-2\gm(t)z, \eqno(1.13)$$ we obtain the following
solution of Navier-Stokes equations (see Theorem 2.4):
$$u=\frac{\al'}{2\al} x+\frac{\be y}{x^2+y^2}+y\sum_{i=0}^\infty\frac{(\al\ptl_t)^i(\Im)}{i!(i+1)!\al}
\left(\frac{x^2+y^2}{4\nu\al}\right)^i,\eqno(1.14)$$
$$v=\frac{\al'}{2\al}y-\frac{\be x}{x^2+y^2}-x\sum_{i=0}^\infty\frac{(\al\ptl_t)^i(\Im)}{i!(i+1)!\al}
\left(\frac{x^2+y^2}{4\nu\al}\right)^i,\eqno(1.15)$$
$$w=\al\sum_{s=0}^\infty\frac{(\al\ptl_t)^s(\vf)}{(s!)^2}
\left(\frac{x^2+y^2}{4\nu\al}\right)^s-\frac{\al'}{\al}z,\eqno(1.16)$$
\begin{eqnarray*}\hspace{2cm}p&=&\frac{\rho((\al')^2-2\al{\al'}')(x^2+y^2)}{8\al^2}
+\frac{\rho(\al{\al'}'-2(\al')^2)z^2}{\al^2}+\be'\arctan\frac{y}{x}
\\ & &+2\nu\rho\al\sum_{i,s=0}^\infty\frac{[(\al\ptl_t)^i(\Im)]
[(\al\ptl_t)^s(\vf)]}{i!(i+1)!s!(s+1)!(\al)^2}
\left(\frac{x^2+y^2}{4\nu\al}\right)^{i+s+1},\hspace{2.4cm}(1.17)
\end{eqnarray*}
where  $\al,\be$ are any functions in $t$ and  $\Im,\vf$ are
arbitrary polynomials in $t$. The above solution can be used to
describe incompressible fluid in a nozzle. The polynomials $\Im$
and $\vf$ can be replaced by the other functions as long as the
related power series converge.

As we emphasized earlier, people are interested in solutions that
are not rotation free. To better capture the rotating nature of
fluid, we introduce the following moving frames:
$$\X=x\cos\al+(y\cos\be+z\sin\be)\sin\al,\;\;
\Y=-x\sin\al+(y\cos\be+z\sin\be)\cos\al,\eqno(1.18)$$
$$\Z=-y\sin\be+z\cos\be,\qquad\U=u\cos\al+(v\cos\be+w\sin\be)
\sin\al,\eqno(1.19)$$
$$\V=-u\sin\al+(v\cos\be+w\sin\be)\cos\al,\;\;
\W=-v\sin\be+w\cos\be,\eqno(1.20)$$ where $\al$ and $\be$ are
functions in $t$. Here we exclude the translation components
because Navier-Stokes equations are invariant  under the
transformations of the type $T_{1\al}$ in (1.8) and (1.9), and we
want to consider solutions modulo these transformations. With
respect to the above rotating frames, Navier-Stokes equations
change to more complicated system of partial differential
equations. Imposing asymmetric conditions on the moving frames, we
find another five families of non-steady rotating solutions with
various parameters. For instance, we have the following solution
of Navier-Stokes equations (see Theorem 3.3):
$$u=\left(\frac{{\al'}'}{2\al'}+6\nu\Y\X^{-2}\right)(\Y\sin\al-\X\cos\al)
-\al'(\X\sin\al+\Y\cos\al),\eqno(1.21)$$
\begin{eqnarray*}\hspace{0.6cm}v&=&-\left(\frac{{\al'}'}
{2\al'}+6\nu\Y\X^{-2}\right)(\X\sin\al+\Y\cos\al)\cos\be
+\al'(\X\cos\al-\Y\sin\al)\cos\be\\ &&-\be'\Z\cos\be
+\left(\be'\X\sin\al+\be'\Y\cos\al-\frac{{\al'}'}{\al'}\Z\right)\sin\be,
\hspace{3.7cm}(1.22)\end{eqnarray*}
\begin{eqnarray*}\hspace{0.6cm}w&=&-\left(\frac{{\al'}'}
{2\al'}+6\nu\Y\X^{-2}\right)(\X\sin\al+\Y\cos\al)\sin\be
+\al'(\X\cos\al-\Y\sin\al)\sin\be\\ &&-\be'\Z\sin\be
+\left(\frac{{\al'}'}{\al'}\Z-\be'\X\sin\al-\be'\Y\cos\al\right)\cos\be,
\hspace{3.6cm}(1.23)\end{eqnarray*}
\begin{eqnarray*}p&=&\rho\{
\frac{(2\al'{{\al'}'}'+4(\al')^4-3({\al'}')^2)(\X^2+\Y^2)}{8(\al')^2}-
\frac{3(\be')^2(\X^2\sin^2\al+\Y^2\cos^2\al)}{2}
\\ & &+12\nu(\al'\Y\X^{-1}-\nu\X^{-2})
+({\be'}'-4\be'\gm)\Z(\X\sin\al+\Y\cos\al)\\ &
&-3(\be')^2\X\Y\sin\al\;\cos \al+
\frac{(\al'{{\al'}'}'+(\al')^2(\be')^2-2({\al'}')^2)\Z^2}{2(\al')^2}
\}.\hspace{3cm}(1.24)\end{eqnarray*}
 The above solution blows up at any
 point on the following rotating plane with a line deleted:
 \begin{eqnarray*}\hspace{2cm}& &\{(x,y,z)\in\mbb{R}^3\mid x\cos\al+y\sin\al\;\cos\be
 +z\sin\al\;\sin\be=0,\\ & &-x\sin\al+y\cos\al\;\cos\be+z\cos\al\;\sin\be\neq
 0\}.\hspace{4.3cm}(1.25)\end{eqnarray*}
This type of solutions may be applied in studying turbulence.
Since all of our solutions in this paper only involve elementary
functions and integrations, they may be applied to engineering
problems with the help of computer, although they appear
sophisticated in format. They can also be used to solve certain
initial value problems for Navier-Stokes equations because they
contain parameter functions.

As we all know that in general,  it is impossible to find all the
solutions of nonlinear partial differential equations both
analytically and algebraically. In our arguments throughout this
paper, we always search for reasonable sufficient conditions of
obtaining exact solutions. For instance, we treat nonzero
functions like ``nonzero constants'' for this purpose because our
approaches in this paper are completely algebraic. Of course, one
can use our methods in this paper to get more solutions, in
particular, by considering the support and discontinuity of the
related functions. We want to remind the reader that we always put
arguments (proofs) before our theorems (conclusions) due to our
purpose of finding exact solutions.

The paper is organized as follows.  Section 2 is devoted to our
asymmetric approaches. We present the general settings for the
moving-frame approach in Section 3 and find two families of exact
solutions. In Section 4, we use the moving frames and certain
ansatzes involving given irrational functions to find another
three families of  exact solutions.

\section{Asymmetric Approaches}

In this section, we will solve incompressible Navier-Stokes
equations (1.1)-(1.4) by imposing asymmetric assumptions on
$u,\;v$ and $w$.

For convenience of computation, we denote
$$\Phi_1=u_t+uu_x+vu_y+wu_z-\nu
(u_{xx}+u_{yy}+u_{zz}),\eqno(2.1)$$
$$\Phi_2=v_t+uv_x+vv_y+wv_z-\nu
(v_{xx}+v_{yy}+v_{zz}),\eqno(2.2)$$
$$\Phi_3=w_t+uw_x+vw_y+ww_z-\nu (w_{xx}+w_{yy}+w_{zz})
.\eqno(2.3)$$ Then Navier-Stokes equations become
 $$\Phi_1+\frac{1}{\rho}p_x=0,\qquad
 \Phi_2+\frac{1}{\rho}p_y=0,\qquad
\Phi_3+\frac{1}{\rho}p_z=0
 \eqno(2.4)$$
 and $u_x+v_y+w_z=0.$ Our strategy is first to solve the following
 compatibility conditions:
 $$\ptl_y(\Phi_1)=\ptl_x(\Phi_2),\qquad
 \ptl_z(\Phi_1)=\ptl_x(\Phi_3),\qquad\ptl_z(\Phi_2)=\ptl_y(\Phi_3)
 \eqno(2.5)$$
and then find $p$ via (2.4).

Let us first look for simplest non-steady solutions of
Navier-Stokes equations (indeed, the corresponding Euler
equations) that are not rotation free. This will help the reader
to better understand our later approaches.  Assume
$$u=\gm_1x-\al_1y-\al_2z,\;\;v=\al_1x+\gm_2y-\al_3z,\;\;w=\al_2x+\al_3y
+\gm_3z,\eqno(2.6)$$ where $\al_i$ and $\gm_i$ are functions in
$t$ such that $\gm_1+\gm_2+\gm_3=0$.  Then
$$\Phi_1=(\gm_1'+\gm_1^2-\al_1^2-\al_2^2)x-(\al_1'-\al_1\gm_3
+\al_2\al_3)y+(\al_1\al_3-\al_2'+\al_2\gm_2)z,\eqno(2.7)$$
$$\Phi_2=(\al_1'-\al_1\gm_3-\al_2\al_3)x+(\gm_2'+\gm_2^2-\al_1^2
-\al_3^2)y-(\al_3'+\al_1\al_2-\al_3\gm_1)z,\eqno(2.8)$$
$$\Phi_3=(\al_2'+\al_1\al_3-\al_2\gm_2)x
+(\al_3'-\al_1\al_2-\al_3\gm_1)y+(\gm_3'+\gm_3^2-\al_2^2
-\al_3^2)z.\eqno(2.9)$$ Furthermore,
$$\ptl_y(\Phi_1)=\ptl_x(\Phi_2)\lra\gm_3=\frac{\al_1'}{\al_1},\eqno(2.10)$$
$$\ptl_z(\Phi_1)=\ptl_x(\Phi_3)\lra
\gm_2=\frac{\al_2'}{\al_2},\eqno(2.11)$$
$$\ptl_z(\Phi_2)=\ptl_y(\Phi_3)\lra
\gm_1=\frac{\al_3'}{\al_3}.\eqno(2.12)$$

Note
$$\gm_1+\gm_2+\gm_3=0\sim \frac{\al_1'}{\al_1}+
\frac{\al_2'}{\al_2}+\frac{\al_3'}{\al_3}=0\sim
\al_1\al_2\al_3=c\eqno(2.13)$$ for some real constant. Moreover,
$$\Phi_1=({\al_3'}'\al_3^{-1}
-\al_1^2-\al_2^2)x-\al_2\al_3y+\al_1\al_3z,\eqno(2.14)$$
$$\Phi_2=-\al_2\al_3x+({\al_2'}'\al_2^{-1}-\al_1^2
-\al_3^2)y-\al_1\al_2z,\eqno(2.15)$$
$$\Phi_3=\al_1\al_3x
-\al_1\al_2y+({\al_1'}'\al^{-1}_1-\al_2^2 -\al_3^2)z.\eqno(2.16)$$
By (2.4),
\begin{eqnarray*}p&=&\frac{\rho}{2}[
(\al_1^2+\al_2^2-{\al_3'}'\al_3^{-1})x^2+
(\al_1^2+\al_3^2-{\al_2'}'\al_2^{-1})y^2+
(\al_2^2+\al_3^2-{\al_1'}'\al_1^{-1})z^2]\\ &
&+\rho(\al_2\al_3xy-\al_1\al_3xz+\al_1\al_2yz)\hspace{7.9cm}(2.17)\end{eqnarray*}
modulo the transformation in (1.10).\psp

{\bf Proposition 2.1}. {\it Let $\al_1,\;\al_2$ and $\al_3$ be
functions in $t$ such that $\al_1\al_2\al_3=c$ for some real
constant $c$. Then we have the following solution of Navier-Stokes
equations (1.1)-(1.4):
$$u=\frac{{\al_3}'}{\al_3}x-\al_1y-\al_2z,\;\;v=\al_1x
+\frac{{\al_2}'}{\al_2}y -\al_3z,\;\;w=\al_2x+\al_3y
+\frac{{\al_1}'}{\al_1}z\eqno(2.18)$$ and $p$ is given in
(2.17).}\psp

 Next we assume
$$v=-\frac{{\be'}'}{2\be'} y,\qquad w=\psi(t,z),\eqno(2.19)$$
where $\be$ is a function in $t$, $\psi$ is a function of $t,z$
and $v$ is so written just for computational convenience by our
earlier experience in [X1, X2]. According to (1.4),
$$u=f(t,y,z)+\left(\frac{{\be'}'}{2\be'}-\psi_z\right)x\eqno(2.20)$$
for some function $f$ of $t,y,z$. Then
\begin{eqnarray*}\hspace{1cm}\Phi_1&=&
f_t+f\left(\frac{{\be'}'}{2\be'}-\psi_z\right)-\frac{{\be'}'}{2\be'}
y f_y+\psi f_z-\nu(f_{yy}+f_{zz})
\\ & &+\left[\left(\frac{{\be'}'}{2\be'}-\psi_z\right)^2
+\frac{\be'{{\be'}'}'-({\be'}')^2}{2(\be')^2}
-\psi_{zt}-\psi\psi_{zz}+\nu\psi_{zzz}\right]x,
\hspace{1.8cm}(2.21)\end{eqnarray*}
$$\Phi_2=\frac{(3({\be'}')^2-2\be'{{\be'}'}')y}{4(\be')^2},\qquad
\Phi_3=\psi_t+\psi\psi_z-\nu\psi_{zz}.\eqno(2.22)$$ Thus (2.5) is
equivalent to the following equations:
$${\cal T}\left[f_t+f\left(\frac{{\be'}'}{2\be'}-\psi_z\right)-\frac{{\be'}'}{2\be'}
y f_y+\psi f_z-\nu(f_{yy}+f_{zz})\right]=0,\eqno(2.23)$$
$${\cal T}\left[\psi_z^2-\frac{{\be'}'}{\be'}\psi_z
-\psi_{zt}-\psi\psi_{zz}+\nu\psi_{zzz}\right]=0\eqno(2.24)$$ with
${\cal T}=\ptl_y,\;\ptl_z$.

Given $b,c\in\mbb{R}$ and a function $\gm$ of $t$, we set
$$\xi_0=be^{\sqrt{\gm'}z+\nu\gm}-ce^{-\sqrt{\gm'}z-\nu\gm},\qquad\xi_1=
b\sin(\sqrt{\gm'}z-\nu\gm),\eqno(2.25)$$
$$\zeta_0=be^{\sqrt{\gm'}z+\nu\gm}+ce^{-\sqrt{\gm'}z-\nu\gm},\qquad\zeta_1=
b\cos(\sqrt{\gm'}z-\nu\gm).\eqno(2.26)$$ Then we have the
following solution of (2.24):
$$\psi=\frac{\xi_r}{\be'\sqrt{(\gm')^3}}-\frac{{\gm'}'}{2\gm'}z\eqno(2.27)$$
with $r=0,1$. Write
$$f=\frac{\hat f}{\sqrt{\be'\gm'}}.\eqno(2.28)$$
The equation (2.23) is implied by the following equation:
$$\hat f_t-\frac{\hat f\zeta_r}{\be'\gm'}-\frac{{\be'}'}{2\be'}
y \hat f_y-\frac{{\gm'}'}{2\gm'}z\hat f_z+ \frac{\hat
f_z\xi_r}{\be'\sqrt{(\gm')^3}}-\nu(\hat f_{yy}+\hat
f_{zz})=0.\eqno(2.29)$$

To solve the above equation, we  assume
$$\hat f=h(t,y)+g(t,y)\zeta_r,\eqno(2.30)$$
where $h$ and $g$ are functions in $t,y$. Then (2.29) is implied
by the following two equations:
$$g_t-\frac{{\be'}'}{2\be'}yg_y-\nu g_{yy}-\frac{h}{\be'\gm'}=0,\eqno(2.31)$$
$$h_t-\frac{{\be'}'}{2\be'}yh_y-\nu h_{yy}
-\frac{(4\dlt_{0,r}bc+\dlt_{1,r}b^2)g}{\be'\gm'}=0.\eqno(2.32)$$
We will solve the above system of partial differential equations
according to the following two cases:\psp

{\it Case 1}. $r=0,\;bc=0$ or $r=1,\;b=0$\psp

In this case,  we have the following solutions:
$$g=\sum_{i=1}^md_{1,i}e^{\nu(a_{1,i}^2-b_{1,i}^2)\be+a_{1,i}
\sqrt{\be'}y}\sin(b_{1,i}(2\nu
a_{1,i}\be+\sqrt{\be'}y)+c_{1,i})+h\int\frac{dt}{\be'\gm'},
\eqno(2.33)$$ $$h=
\sum_{s=1}^nd_{2,s}e^{\nu(a_{2,s}^2-b_{2,s}^2)\be +a_{2,s}
\sqrt{\be'}y}\sin(b_{2,s}(2\nu
a_{2,s}\be+\sqrt{\be'}y)+c_{2,s}),\eqno(2.34)$$ where
$a_{1,i},b_{1,i},c_{1,i},d_{1,i}$ and
$a_{2,s},b_{2,s},c_{2,s},d_{2,s}$ are real constants. \psp

{\it Case 2}. $r=0,\;bc\neq 0$ or $r=1,\;b\neq 0$ if $r=1$.\psp

For convenience, we denote
$$a=\sqrt{4\dlt_{0,r}bc+\dlt_{1,r}b^2}.\eqno(2.35)$$
Then we have the following solution of the system (2.31) and
(2.32):
\begin{eqnarray*}& &g=\cosh\left(a\int\frac{dt}{\be'\gm'}\right)
\sum_{i=1}^md_{1,i}e^{\nu(a_{1,i}^2-b_{1,i}^2)\be+a_{1,i}
\sqrt{\be'}y}\sin(b_{1,i}(2\nu a_{1,i}\be+\sqrt{\be'}y)+c_{1,i})\\
& &+\frac{
\sinh\left(a\int\frac{dt}{\be'\gm'}\right)}{a}\sum_{s=1}^nd_{2,s}e^{\nu(a_{2,s}^2-b_{2,s}^2)\be
+a_{2,s} \sqrt{\be'}y}\sin(b_{2,s}(2\nu
a_{2,s}\be+\sqrt{\be'}y)+c_{2,s}),
\hspace{0.5cm}(2.36)\end{eqnarray*}
\begin{eqnarray*}& &h=a\sinh\left(a\int\frac{dt}{\be'\gm'}\right)
\sum_{i=1}^md_{1,i}e^{\nu(a_{1,i}^2-b_{1,i}^2)\be+a_{1,i}
\sqrt{\be'}y}\sin(b_{1,i}(2\nu a_{1,i}\be+\sqrt{\be'}y)+c_{1,i})\\
& &+ \cosh\left(\int\frac{adt}{\be'\gm'}\right)
\sum_{s=1}^nd_{2,s}e^{\nu(a_{2,s}^2-b_{2,s}^2)\be +a_{2,s}
\sqrt{\be'}y}\sin(b_{2,s}(2\nu a_{2,s}\be+\sqrt{\be'}y)+c_{2,s}).
\hspace{0.4cm}(2.37)\end{eqnarray*} \pse

In any case,
$$\Phi_1=\left(\frac{2\be'{{\be'}'}'-({\be'}')^2}{4(\be')^2}
+\frac{2\gm'{{\gm'}'}'-({\gm'}')^2}{4(\gm')^2}+\frac{{\be'}'{\gm'}'}
{2\be'\gm'}+\frac{\dlt_{0,r}4bc+\dlt_{1,r}b^2}{(\be'\gm')^2}\right)x
,\eqno(2.38)$$
$$\Phi_3=\left[\frac{(3({\gm'}')^2-2\gm'{{\gm'}'}')z^2}
{8(\gm')^2}+\frac{\xi_r^2-(-1)^r(2{\be'}'\gm'+5\be'{\gm'}')
\zeta_r}{2(\be')^3(\gm')^3}\right]_z. \eqno(2.39)$$ Thus (2.4),
(2.38), (3.39) and the first equation in (2.22) give
\begin{eqnarray*}p&=&
\rho\left[\frac{(2\be'{{\be'}'}'-3({\be'}')^2)y^2}{8(\be')^2}
+\frac{2\gm'{{\gm'}'}'-3({\gm'}')^2}
{8(\gm')^2}z^2+\frac{(-1)^r(2{\be'}'\gm'+5\be'{\gm'}')
\zeta_r-\xi_r^2}{2(\be')^3(\gm')^3}\right]\\ & &+\frac{\rho
x^2}{2} \left(\frac{({\be'}')^2-2\be'{{\be'}'}'}{4(\be')^2}
+\frac{({\gm'}')^2-2\gm'{{\gm'}'}'}{4(\gm')^2}-\frac{{\be'}'{\gm'}'}
{2\be'\gm'}-\frac{\dlt_{0,r}4bc+\dlt_{1,r}b^2}{(\be'\gm')^2}\right)
\hspace{1.3cm}(2.40)\end{eqnarray*} modulo the transformation in
(1.10).

By (2.20), (2.27), (2.28), (2.33), (2.34), (2.36) and (2.37), we
obtain:\psp

{\bf Theorem 2.2}. {\it Let $\be,\gm$ be functions in $t$, and let
$b,c,a_{1,i},b_{1,i},c_{1,i},d_{1,i}$ and
$a_{2,s},b_{2,s},c_{2,s},\\ d_{2,s}$ be real constants. Define
$\xi_r$ and $\zeta_r$ in (2.25) and (2.26). For $r=0,1$,  we have
the following solution of Navier-Stokes equations (1.1)-(1.4):
$$v=-\frac{{\be'}'}{2\be'} y,\qquad w=
\frac{\xi_r}{\be'\sqrt{(\gm')^3}}-\frac{{\gm'}'}{2\gm'}z,\eqno(2.41)$$
$p$ is given (2.40), and
\begin{eqnarray*}\hspace{1cm}u&=&\frac{\zeta_r}{\sqrt{\be'\gm'}}
\sum_{i=1}^md_{1,i}e^{\nu(a_{1,i}^2-b_{1,i}^2)\be+a_{1,i}
\sqrt{\be'}y}\sin(b_{1,i}(2\nu a_{1,i}\be+\sqrt{\be'}y)+c_{1,i})\\
& &+\left(\frac{{\be'}'}{2\be'}
+\frac{{\gm'}'}{2\gm'}-\frac{\zeta_r}{\be'\gm'}\right)x
+\frac{1+\zeta_r\int\frac{dt}{\be'\gm'}}{\sqrt{\be'\gm'}}
\sum_{s=1}^nd_{2,s}e^{\nu(a_{2,s}^2-b_{2,s}^2)\be +a_{2,s}
\sqrt{\be'}y}\\ & &\times\sin(b_{2,s}(2\nu
a_{2,s}\be+\sqrt{\be'}y)+c_{2,s})\hspace{6.6cm}
(2.42)\end{eqnarray*} if $r=0,\;bc=0$, and
\begin{eqnarray*}u&=&\left(\frac{{\be'}'}{2\be'}+\frac{{\gm'}'}{2\gm'}-
\frac{\zeta_r}{\be'\gm'}\right)x+
\frac{1}{\sqrt{\be'\gm'}}\left(\zeta_r\cosh
\left(\int\frac{adt}{\be'\gm'}\right)
+a\sinh\left(\int\frac{adt}{\be'\gm'}\right)\right)
\\ & &\times\sum_{i=1}^md_{1,i}e^{\nu(a_{1,i}^2-b_{1,i}^2)\be+a_{1,i}
\sqrt{\be'}y}\sin(b_{1,i}(2\nu a_{1,i}\be+\sqrt{\be'}y)+c_{1,i})\\
& & + \frac{1}{\sqrt{\be'\gm'}}
\left(\cosh\left(\int\frac{adt}{\be'\gm'}\right)
+\frac{\zeta_r}{a}\sinh\left(\int\frac{adt}{\be'\gm'}\right)\right)
\sum_{s=1}^nd_{2,s}e^{\nu(a_{2,s}^2-b_{2,s}^2)\be +a_{2,s}
\sqrt{\be'}y}\\ & &\times\sin(b_{2,s}(2\nu
a_{2,s}\be+\sqrt{\be'}y)+c_{2,s})\hspace{7.6cm}
(2.43)\end{eqnarray*} if $r=0,\;bc\neq 0$ or $r=1,\;b\neq 0$,
where $a$ is defined in (2.35).} \psp

{\bf Remark 2.3}. We can use Fourier expansion to solve the system
(2.31) and (2.32) for $g(\be,\sqrt{\be'}y)$ and
$h(\be,\sqrt{\be'}y)$ with given $g(0,\sqrt{\be'}y)$ and
$h(0,\sqrt{\be'}y)$. In this way, we can obtain discontinuous
solutions of Navier-Stokes equations (1.1)-(1.4), which may be
useful in studying shock waves.\psp

Set
$$\varpi=x^2+y^2.\eqno(2.44)$$
Consider
$$u=\frac{\al'}{2\al} x+y\phi(t,\varpi),\qquad v=\frac{\al'}{2\al}y
-x\phi(t,\varpi),\qquad
w=\psi(t,\varpi)-\frac{\al'}{\al}z,\eqno(2.45)$$ where $\al$ is a
function in $t$ and $\phi,\psi$ are functions in $t,\varpi$. Then
(2.1)-(2.3) give
$$\Phi_1=\frac{2\al{\al'}'-(\al')^2}{4\al^2} x+y\phi_t
+\frac{\al'y}{\al}(\varpi\phi)_\varpi-x\phi^2-
4y\nu(\varpi\phi)_{\varpi\varpi},\eqno(2.46)$$
$$\Phi_2=\frac{2\al{\al'}'-(\al')^2}{4\al^2}y-x\phi_t
-\frac{\al'x}{\al}(\varpi\phi)_\varpi-y\phi^2+
4x\nu(\varpi\phi)_{\varpi\varpi},\eqno(2.47)$$
$$\Phi_3=\frac{2(\al')^2-\al{\al'}'}{\al^2}z+\psi_t
-\frac{\al'}{\al}\psi+\frac{\al'}{\al}\varpi
\psi_\varpi-4\nu(\psi_\varpi+\varpi\psi_{\varpi\varpi}).\eqno(2.48)$$

Note that $\ptl_y(\Phi_1)=\ptl_x(\Phi_2)$ becomes
$$(\varpi\phi)_{\varpi t}+\frac{\al'}{\al}
((\varpi\phi)_\varpi+\varpi(\varpi\phi)_{\varpi\varpi})
-4\nu((\varpi\phi)_{\varpi\varpi}+
\varpi(\varpi\phi)_{\varpi\varpi\varpi})=0.\eqno(2.49)$$ Set
$$\hat\phi=\al(\varpi\phi)_\varpi.\eqno(2.50)$$
Then (2.49) becomes
$$\hat\phi_t+\frac{\al'}{\al}\varpi\hat\phi_\varpi-
4\nu(\hat\phi_\varpi+\varpi\hat\phi_{\varpi\varpi})=0.\eqno(2.51)$$
So we have the solution
$$\hat\phi=\sum_{i=0}^\infty\frac{(\al\ptl_t)^i(\Im)}{(i!)^2}
\left(\frac{\varpi}{4\nu\al}\right)^i\eqno(2.52)$$ for a
polynomial $\Im$ in $t$. By (2.50)-(2.52), we have
$$\phi=\be\varpi^{-1}+\sum_{i=0}^\infty\frac{(\al\ptl_t)^i(\Im)}{i!(i+1)!\al}
\left(\frac{\varpi}{4\nu\al}\right)^i\eqno(2.53)$$ for a function
$\be$ of $t$.

Note
$$\phi_t=\be'\varpi^{-1}+\sum_{i=0}^\infty\left[\frac{(\al\ptl_t)^{i+1}(\Im)}{i!(i+1)!\al^2}
-\frac{\al'(\al\ptl_t)^i(\Im)}{(i!)^2!\al^2}\right]
\left(\frac{\varpi}{4\nu\al}\right)^i,\eqno(2.54)$$
$$\frac{\al'}{\al}
((\varpi\phi)_\varpi=\frac{\al'}{\al^2}\hat\phi=
\frac{\al'}{\al^2}\sum_{i=0}^\infty\frac{(\al\ptl_t)^i(\Im)}{(i!)^2}
\left(\frac{\varpi}{4\nu\al}\right)^i,\eqno(2.55)$$
$$4\nu(\varpi\phi)_{\varpi\varpi}=\frac{4\nu\hat\phi_\varpi}{\al}=
\sum_{i=1}^\infty\frac{(\al\ptl_t)^i(\Im)}{(i-1)!i!\al^2}
\left(\frac{\varpi}{4\nu\al}\right)^{i-1}=\sum_{i=0}^\infty\frac{(\al\ptl_t)^{i+1}
(\Im)}{i!(i+1)!\al^2}
\left(\frac{\varpi}{4\nu\al}\right)^i.\eqno(2.56)$$ Thus
$$\phi_t+\frac{\al'}{\al}(\varpi\phi)_\varpi-
4\nu(\varpi\phi)_{\varpi\varpi}=\be'\varpi^{-1}.\eqno(2.57)$$
Therefore,
$$\Phi_1=
\frac{2\al{\al'}'-(\al')^2}{4\al^2}
x+\frac{\be'y}{x^2+y^2}-x\phi^2\eqno(2.58)$$ and
$$\Phi_2=
\frac{2\al{\al'}'-(\al')^2}{4\al^2}
y-\frac{\be'x}{x^2+y^2}-y\phi^2.\eqno(2.59)$$

On the other hand, Equations $\ptl_z(R_1)=\ptl_x(R_3)$ and
$\ptl_z(R_2)=\ptl_y(R_3)$ are implied by the following
differential equation:
$$\psi_t
-\frac{\al'}{\al}\psi+\frac{\al'}{\al}\varpi
\psi_\varpi-4\nu(\psi_\varpi+\varpi\psi_{\varpi\varpi})=0\eqno(2.60)$$
(cf. (2.48)). Similarly, we have the solution:
$$\psi=\al\sum_{s=0}^\infty\frac{(\al\ptl_t)^s(\vf)}{(s!)^2}
\left(\frac{\varpi}{4\nu\al}\right)^s,\eqno(2.61)$$ where $\vf$ is
a polynomial in $t$. With this $\psi$,
$$\Phi_3=\frac{2(\al')^2-\al{\al'}'}{\al^2}z.\eqno(2.62)$$
By (2.4), (2.44), (2.45), (2.53), (2.58), (2.59), (2.61) and
(2.62), we obtain:\psp

{\bf Theorem 2.4}. {\it Let $\al,\be$ be any functions in $t$ and
let $\Im,\vf$ be polynomials in $t$. We have the following
solution of Navier-Stokes equations (1.1)-(1.4):
$$u=\frac{\al'}{2\al} x+\frac{\be y}{x^2+y^2}+y\sum_{i=0}^\infty\frac{(\al\ptl_t)^i(\Im)}{i!(i+1)!\al}
\left(\frac{x^2+y^2}{4\nu\al}\right)^i,\eqno(2.63)$$
$$v=\frac{\al'}{2\al}y-\frac{\be x}{x^2+y^2}-x\sum_{i=0}^\infty\frac{(\al\ptl_t)^i(\Im)}{i!(i+1)!\al}
\left(\frac{x^2+y^2}{4\nu\al}\right)^i,\eqno(2.64)$$
$$w=\al\sum_{s=0}^\infty\frac{(\al\ptl_t)^s(\vf)}{(s!)^2}
\left(\frac{x^2+y^2}{4\nu\al}\right)^s-\frac{\al'}{\al}z,\eqno(2.65)$$
\begin{eqnarray*}\hspace{2cm}p&=&\frac{\rho((\al')^2-2\al{\al'}')(x^2+y^2)}{8\al^2}
+\frac{\rho(\al{\al'}'-2(\al')^2)z^2}{\al^2}+\be'\arctan\frac{y}{x}
\\ & &+2\nu\rho\al\sum_{i,s=0}^\infty\frac{[(\al\ptl_t)^i(\Im)]
[(\al\ptl_t)^s(\vf)]}{i!(i+1)!s!(s+1)!(\al)^2}
\left(\frac{x^2+y^2}{4\nu\al}\right)^{i+s+1}.\hspace{2.4cm}(2.66)
\end{eqnarray*}}
\pse

{\bf Remark 2.5}. The above solution can be used to describe
incompressible fluid in a nozzle. The polynomials $\Im$ and $\vf$
can be replaced by the other functions as long as the related
power series converge.

\section{Moving-Frame Approach I}

In this section, we will present the general settings for the
moving-frame approach and find two families of exact solutions.

Let $\al,\be$ be given functions of $t$. Denote
$$T=\left(\begin{array}{ccc}\cos\al&\sin\al\;\cos\be&\sin\al\;\sin\be\\
-\sin\al&\cos\al\;\cos\be&\cos\al\;\sin\be\\ 0&
-\sin\be&\cos\be\end{array}\right)\eqno(3.1)$$ and
$$Q=\left(\begin{array}{ccc}0&\al'&\be'\sin\al\\
-\al'&0&\be'\cos\al\\-\be'\sin\al&-\be'\cos\al&0
\end{array}\right).\eqno(3.2)$$
Then
$$T^{-1}=T^t=\left(\begin{array}{ccc}\cos\al&-\sin\al&0\\
\sin\al\;\cos\be &\cos\al\;\cos\be&-\sin\be
\\\sin\al\;\sin\be&\cos\al\;\sin\be&\cos\be\end{array}\right)\eqno(3.3)$$
and
$$\frac{d}{dt}(T)=QT.\eqno(3.4)$$
Define the moving frames:
$$\left(\begin{array}{c}{\cal U}\\{\cal V}\\ {\cal
W}\end{array}\right)=T\left(\begin{array}{c}u\\ v\\
w\end{array}\right),\qquad \left(\begin{array}{c}{\cal X}\\{\cal
Y}\\ {\cal
Z}\end{array}\right)=T\left(\begin{array}{c}x\\ y\\
z\end{array}\right).\eqno(3.5)$$ Note
$$\Dlt=\ptl_x^2+\ptl_y^2+\ptl_z^2=\ptl_{\cal X}^2+\ptl_{\cal Y}^2+
\ptl_{\cal Z}^2,\qquad
u\ptl_x+v\ptl_y+z\ptl_z=\U\ptl_\X+\V\ptl_\Y+\W\ptl_\Z,\eqno(3.6)$$
$$u_x+v_y+w_z=\U_\X+\V_\Y+\W_\Z,\qquad
 \left(\begin{array}{c}\X_t\\ \Y_t\\ Z_t\end{array}\right)=
 Q\left(\begin{array}{c}\X\\\Y\\ \Z\end{array}\right),\eqno(3.7)$$
$$\left(\begin{array}{c}\ptl_\X\\
\ptl_\Y\\ \ptl_\Z\end{array}\right)=T\left(\begin{array}{c}\ptl_x\\
\ptl_y\\ \ptl_z\end{array}\right),\qquad
\left(\begin{array}{c}\ptl_t(\U)\\ \ptl_t(\V)\\
\ptl_t(\W)\end{array}\right)=Q\left(\begin{array}{c}\U\\ \V\\
\W\end{array}\right)+T\left(\begin{array}{c}u_t\\ v_t\\
w_t\end{array}\right).\eqno(3.8)$$

Write $\U,\V,\W,p$ as functions in $t,\X,\Y,\Z$. Set
\begin{eqnarray*}& &R_1=\U_t+\al'(\Y\U_\X-\X\U_\Y
-\V)+\be'(\Z\U_\X-\X\U_\Z-\W)\sin\al\\
&
&+\be'(\Z\U_\Y-\Y\U_\Z)\cos\al+\U\U_\X+\V\U_\Y+\W\U_\Z-\nu\Dlt(\U),
\hspace{4cm}(3.9)\end{eqnarray*}
\begin{eqnarray*}& &R_2=\V_t+\al'(\Y\V_\X-\X\V_\Y
+\U)+\be'(\Z\V_\X-\X\V_\Z)\sin\al\\
&&+\be'(\Z\V_\Y-\Y\V_\Z-\W)\cos\al+\U\V_\X+\V\V_\Y+\W\V_\Z-\nu\Dlt(\V),
\hspace{3cm}(3.10)\end{eqnarray*}
\begin{eqnarray*}& &R_3=\W_t+\al'(\Y\W_\X-\X\W_\Y
)+\be'(\Z\W_\X-\X\W_\Z+\U)\sin\al\\&&+\be'(\Z\W_\Y-\Y\W_\Z+\V)\cos\al
+\U\W_\X+\V\W_\Y+\W\W_\Z-\nu\Dlt(\W),\hspace{2.3cm}(3.11)\end{eqnarray*}
Then Navier-Stokes equations (1.1)-(1.4) become
$$R_1+\frac{1}{\rho}p_{_\X}=0,\qquad R_2+
\frac{1}{\rho}p_{_\Y}=0,\qquad R_3+
\frac{1}{\rho}p_{_\Z}=0,\eqno(3.12)$$
$$\U_\X+\V_\Y+\W_\Z=0.\eqno(3.13)$$
Instead of solving the equations in (3.12), we will first solve
the following compatibility equations:
$$\ptl_\Y(R_1)=\ptl_\X(R_2),\qquad\ptl_\Z(R_1)=\ptl_\X(R_3),
\qquad \ptl_\Z(R_2)=\ptl_\Y(R_3)\eqno(3.14)$$ for $\U,\V,\W$, and
then find $p$ from the equations in (3.12).

Let $f$ be a function in $t,\Y,\Z$ such that
$\ptl_\Y^2(f)=\ptl_\Z^2(f)=0$, and let $\phi,\psi$ be functions in
$t,\X$. Suppose that $\gm$ is a function of $t$. Assume
$$\U=f-2\gm'\X,\qquad\V=\phi+\gm'\Y,\qquad\W=\psi+\gm'\Z
.\eqno(3.15)$$ Then
\begin{eqnarray*}\hspace{0.5cm}& &R_1=f_t-2{\gm'}'\X-\al'(3\gm'\Y+\X f_\Y+\phi
)-\be'(3\gm'\Z+\X f_\Z+\psi)\sin\al \\ & &+\be'(\Z f_\Y-\Y
f_\Z)\cos\al-2\gm'(f-2\gm'\X)+f_\Y(\phi+\gm'\Y)+f_\Z(\psi+\gm'\Z)
,\hspace{1cm}(3.16)\end{eqnarray*}
\begin{eqnarray*}\hspace{1.3cm}R_2&=&\phi_t+{\gm'}'\Y+\al'(\Y\phi_\X-3\gm'\X +f)
+\be'\Z\phi_\X\sin\al-\be'\psi\cos\al\\
&&+(f-2\gm'\X)\phi_\X+\gm'\phi+(\gm')^2\Y
-\nu\phi_{\X\X},\hspace{4.8cm}(3.17)\end{eqnarray*}
\begin{eqnarray*}\hspace{1cm}R_3&=&\psi_t+{\gm'}'\Z+\al'\Y\psi_\X
+\be'(\Z\psi_\X-3\gm'\X+f)\sin\al-\nu\psi_{\X\X}\\&&
+\be'\phi\cos\al
+(f-2\gm'\X)\psi_\X+\gm'(\psi+\gm'\Z).\hspace{4.7cm}(3.18)\end{eqnarray*}
Now (3.14) becomes
\begin{eqnarray*}\hspace{1.5cm}& &
\phi_{t\X}+(\al'\Y
+\be'\Z\sin\al+f)\phi_{\X\X}-\be'\psi_\X\cos\al-2\gm'(\X\phi_\X)_\X
\\ & &+\gm'\phi_\X -\nu\phi_{\X\X\X}=f_{t\Y}-\be'
f_\Z\cos\al-\gm'f_\Y,\hspace{4.8cm}(3.19)\end{eqnarray*}
\begin{eqnarray*}\hspace{1.5cm}& &\psi_{t\X}+(\al'\Y
+\be'\Z\sin\al+f)\psi_{\X\X}-\nu\psi_{\X\X\X} +\be'\phi_\X\cos\al
\\&&-2(\gm'\X\psi_\X)_\X+\gm'\psi_\X=
f_{t\Z}+\be'
f_\Y\cos\al-\gm'f_\Z,\hspace{4cm}(3.20)\end{eqnarray*}
$$\al'f_\Z +(\be'\sin\al+f_\Z)\phi_\X=
(\al'+f_\Y)\psi_\X +\be'f_\Y\sin\al.\eqno(3.21)$$

By (3.19) and (3.20), we take
$$f=-\al'\Y-\be'\Z\sin\al\eqno(3.22)$$
modulo the transformations of type $T_{1\al}$ in (1.8) and (1.9).
Note that (3.21) is implied by (3.22). Integrating (3.19) and
(3.20), we obtain
$$\phi_t-2\gm'\X\phi_\X +\gm'\phi -\nu\phi_{\X\X}-\be'\psi\cos\al
=[(\be')^2
\sin\al\;\cos\al+\al'\gm'-{\al'}']\X+\be_1,\eqno(3.23)$$
\begin{eqnarray*}\hspace{2cm}& &\psi_t-2\gm'\X\psi_\X
+\gm'\psi-\nu\psi_{\X\X} +\be'\phi\cos\al \\ &=& -[(\be'\sin\al)'
+\al'\be'\cos\al-\gm'\be'\sin\al]\X+\be_2,
\hspace{3.9cm}(3.24)\end{eqnarray*} where $\be_1$ and $\be_2$ are
arbitrary functions of $t$. To solve the above problem, we write
$$\be'=\frac{\vf'}{\cos\al},\qquad \gm=\frac{1}{4}\ln\mu'\eqno(3.25)$$
and set
$$\left(\begin{array}{c}\hat\phi\\ \hat\psi\end{array}\right)
=\sqrt[4]{\mu'}\left(\begin{array}{cc}\cos\vf&-\sin\vf\\
\sin\vf&\cos\vf\end{array}\right) \left(\begin{array}{c}\phi\\
\psi\end{array}\right),\eqno(3.26)$$
$$\left(\begin{array}{c}\gm_1\\ \gm_2\end{array}\right)
=\int\frac{1}{\sqrt[4]{\mu'}} \left(\begin{array}{cc}\cos\vf&
\sin\vf\\
-\sin\vf&\cos\vf\end{array}\right) \left(\begin{array}{c}(\vf')^2
\tan\al+\frac{\al'{\mu'}'}{4\mu'}-{\al'}'
\\ -(\vf'\tan\al)'-\al'\vf'+\frac{{\mu'}'\vf'}{4\mu'}\tan\al
\end{array}\right)dt.\eqno(3.27)$$
Then (3.23) and (3.24) are equivalent to:
$$\hat\phi_t-\frac{{\mu'}'}{2\mu'}\X\hat\phi_\X
-\nu\phi_{\X\X}=\gm_1'\sqrt{\mu'}\X+\vf_1',\eqno(3.28)$$
$$\hat\psi_t-\frac{{\mu'}'}{2\mu'}\X\hat\psi_\X
-\nu\psi_{\X\X}=\gm_2'\sqrt{\mu'}\X+\vf_2',\eqno(3.29)$$ where
$\vf_1$ and $\vf_2$ are arbitrary functions of $t$.
 Thus
$$\hat\phi=\gm_1\sqrt{\mu'}\X+\vf_1+\sum_{i=1}^md_{1,i}e^{\nu
(a_{1,i}^2-b_{1,i}^2)\mu+a_{1,i}\sqrt{\mu'}\X}\sin(b_{1,i}(2a_{1,i}
\mu+\sqrt{\mu'}\X)+c_{i,1}),\eqno(3.30)$$
$$\hat\psi=\gm_2\sqrt{\mu'}\X+\vf_2+ \sum_{s=1}^nd_{1,s}e^{\nu
(a_{2,s}^2-b_{1,s}^2)\mu+a_{2,s}\sqrt{\mu'}\X}\sin(b_{2,s}(2a_{2,s}
\mu+\sqrt{\mu'}\X)+c_{2,s}),\eqno(3.31)$$ where
$a_{1,i},a_{2,s},b_{1,i},b_{2,s},c_{1,i},c_{2,s},d_{1,i}$ and
$d_{2,s}$ are real constants. According (3.26), we have
\begin{eqnarray*}\hspace{1cm}\phi&=&\sqrt[4]{\mu'}(\gm_1\cos\vf+\gm_2\sin\vf)
\X+\sgm_1+\frac{\cos\vf}{\sqrt[4]{\mu'}}\sum_{i=1}^md_{1,i}e^{\nu
(a_{1,i}^2-b_{1,i}^2)\mu+a_{1,i}\sqrt{\mu'}\X}\\
& &\times\sin(b_{1,i}(2a_{1,i}+\sqrt{\mu'}\X)+c_{i,1})+
\frac{\sin\vf}{\sqrt[4]{\mu'}}\sum_{s=1}^nd_{1,s}e^{\nu
(a_{2,s}^2-b_{1,s}^2)\mu+a_{2,s}\sqrt{\mu'}\X}\\
& &\times\sin(b_{2,s}(2a_{2,s}+\sqrt{\mu'}\X)+c_{2,s}),
\hspace{6.9cm}(3.32)\end{eqnarray*}
\begin{eqnarray*}\hspace{1cm}\psi&=&\sqrt[4]{\mu'}(\gm_2\cos\vf-\gm_1\sin\vf)
\X+\sgm_2-\frac{\sin\vf}{\sqrt[4]{\mu'}}\sum_{i=1}^md_{1,i}e^{\nu
(a_{1,i}^2-b_{1,i}^2)\mu+a_{1,i}\sqrt{\mu'}\X}\\
& &\times\sin( b_{1,i}(2a_{1,i}+\sqrt{\mu'}\X)+c_{i,1})+
\frac{\cos\vf}{\sqrt[4]{\mu'}}\sum_{s=1}^nd_{1,s}e^{\nu
(a_{2,s}^2-b_{1,s}^2)\mu+a_{2,s}\sqrt{\mu'}\X}\\
& &\times\sin(b_{2,s}(2a_{2,s} +\sqrt{\mu'}\X)+c_{2,s}),
\hspace{6.9cm}(3.33)\end{eqnarray*} where $\sgm_1$ and $\sgm_2$
are arbitrary functions of $t$.

To find the pressure $p$, we recalculate
\begin{eqnarray*}R_1&=&((\vf')^2\Y-2\vf'\psi)\tan\al-2\al'\phi
-{\al'}'\Y-({\vf'}'+\al'\vf'(1+\sec^2\al))\Z
\\ & &-\frac{{\mu'}'(\al'\Y+\vf'\Z\tan\al)}{\mu'}+((\al')^2+(\vf')^2
\tan^2\al) \X+\frac{(3({\mu'}')^2-2{{\mu'}'}')\X}{4(\mu')^2}
,\hspace{0.9cm}(3.34)\end{eqnarray*}
$$R_2=\frac{(4{{\mu'}'}'-3({\mu'}')^2)\Y}{16(\mu')^2}
-(\al')^2\Y+ \sgm_1'+(\vf'\X-\al'\Z)\vf'\tan\al-
\frac{(\al'{\mu'}'+{\al'}'\mu')\X}{\mu'},\eqno(3.35)$$
\begin{eqnarray*}\hspace{2cm}R_3&=&\frac{(4{{\mu'}'}'-3({\mu'}')^2)\Z}{16(\mu')^2}
-\frac{{\mu'}'\X+\al'\mu'\Y}{\mu'}\vf'\tan\al+ \sgm_2'\\&
&-(\vf')^2\Z\tan^2\al-({\vf'}'+\al'\vf'(1+\sec^2\al))\X.
\hspace{3.7cm}(3.36)
\end{eqnarray*}
Thus
\begin{eqnarray*}p&=&\rho\{
({\al'}'\Y+({\vf'}'\tan\al+\al'\vf'(1+\sec^2\al))\Z)\X
+\frac{{\mu'}'(\al'\Y+\vf'\Z\tan\al)\X}{\mu'}\\
& &-\sgm_1'\Y-\sgm_2'\Z+\frac{\X^2}{2}
\left(\frac{({{\mu'}'}'-(3{\mu'}')^2)}{4(\mu')^2}-(\al')^2-(\vf')^2\tan^2\vf\right)
\\ &
&+\frac{(3({\mu'}')^2-4{{\mu'}'}')(\Y^2+\Z^2)}{32(\mu')^2}+(\al'\Z-\vf'\X)\vf'\Y\tan\al
\\ & &+\frac{(\al')^2\Y^2+(\gm')^2\Z^2\tan^2\al}{2}
+2\int(\al'\phi+\vf'\psi\tan\al)d\X\}
\hspace{3.9cm}(3.37)\end{eqnarray*} modulo the transformation in
(1.10). In summary, we have:\psp

{\bf Theorem 3.1}. {\it Let $\al,\vf,\mu,\sgm_1,\sgm_2$ be
functions of $t$ with $\mu'> 0$. Take real constants
$\{a_{1,i},a_{2,s},b_{1,i},b_{2,s},c_{1,i},c_{2,s},d_{1,i},d_{2,s}\mid
i=1,...,m;s=1,...,n\}$. Denote
$$\be=\int\frac{\vf'dt}{\cos\al}\eqno(3.38)$$
and define $\gm_1,\gm_2$ by (3.27). Take the notations $\X,\Y,\Z$
given in (3.1) and (3.5). In terms of the functions $\phi$ in
(3.32) and $\psi$ in (3.33), we have the following solution of the
Navier-Stokes equations (1.1)-(1.4):
$$u=-\left(\frac{{\mu'}'\X}{2\mu'}
+\al'\Y+\vf'\Z\tan\al\right)\cos\al
-\left(\phi+\frac{{\mu'}'\Y}{4\mu'}\right)\sin\al,\eqno(3.39)$$
\begin{eqnarray*}\hspace{2cm}v&=&\left(\frac{{\mu'}'\X}{2\mu'}
-\al'\Y-\vf'\Z\tan\al\right)\sin\al\;\cos\be\\ &&
+\left(\phi+\frac{{\mu'}'\Y}{4\mu'}\right)\cos\al\;\cos\be-
\left(\psi+\frac{{\mu'}'\Z}{4\mu'}\right)\sin\be,
\hspace{3.2cm}(3.40)\end{eqnarray*}
\begin{eqnarray*}\hspace{2cm}w&=&\left(\frac{{\mu'}'\X}{2\mu'}
-\al'\Y-\vf'\Z\tan\al\right)\sin\al\;\sin\be\\ &&
+\left(\phi+\frac{{\mu'}'\Y}{4\mu'}\right)\cos\al\;\sin\be+
\left(\psi+\frac{{\mu'}'\Z}{4\mu'}\right)\cos\be
\hspace{3.3cm}(3.41)\end{eqnarray*} and $p$ is given in
(3.37).}\psp

{\bf Remark 3.2}. We can use Fourier expansion to solve the system
(3.28) and (3.29) for $\hat\phi(\mu,\sqrt{\mu'}\X)$ and
$\hat\psi(\mu,\sqrt{\mu'}\X)$ with given
$\hat\phi(0,\sqrt{\mu'}\X)$ and $\hat\psi(0,\sqrt{\mu'}\X)$. In
this way, we can obtain discontinuous solutions of Navier-Stokes
equations (1.1)-(1.4), which may be useful in studying shock
waves.\psp

Let $f,g,h$ be functions of $t,\X,\Y,\Z$ that are linear in
$\X,\Y,\Z$ and $f_\X+g_{_\Y}+h_\Z=0$. Based on our experience in
[X2], we assume
$$\U=f-6\nu\X^{-1},\qquad\V=g-6\nu\Y\X^{-2},\qquad\W=h.\eqno(3.42)$$
Then
\begin{eqnarray*}\hspace{1.1cm}R_1&=&f_t+ff_\X+f_\Y g+f_\Z h
-6\nu f_\X\X^{-1}+\al'(\Y f_\X-\X f_\Y-g)
\\
& &+\be'(\Z f_\X-\X f_\Z-h)\sin\al
+\be'(\Z f_\Y-\Y f_\Z)\cos\al \\
& &+6\nu (f-\Y
f_\Y+2\al'\Y+\be'\Z\sin\al)\X^{-2}-24\nu^2\X^{-3},\hspace{3.2cm}(3.43)\end{eqnarray*}
\begin{eqnarray*}\hspace{0.5cm}R_2&=&g_t+fg_{_\X}+gg_{_\Y}+g_{_\Z}h+
\al'(\Y g_{_\X}-\X g_{_\Y} +f)+\be'(\Z g_{_\X}-\X g_{_\Z})\sin\al\\
& &-6\nu (\al'+g_{_\X})\X^{-1} -6\nu(g+\be'\Z\cos\al-\al'\X+\Y
g_{_\Y})\X^{-2}\\ & &+\be'(\Z g_{_\Y}-g_{_\Z}\Y-h)\cos\al
+12\nu\Y(f+\al'\Y+\be'\Z\sin\al)\X^{-3},
\hspace{1.8cm}(3.44)\end{eqnarray*}
\begin{eqnarray*}& &R_3=h_t+fh_\X+gh_\Y+hh_\Z
+\al'(\Y h_\X-\X h_\Y )+\be'(\Z h_\X-\X h_\Z+f)\sin\al\\&&+\be'(\Z
h_\Y-\Y h_\Z+g)\cos\al -6\nu(h_x+\be'\sin\al)\X^{-1}
-6\nu(h_\Y+\be'\cos\al) \Y\X^{-2}.
\hspace{0.3cm}(3.45)\end{eqnarray*} According the negative powers
of $\X$ in (3.14), we take
$$f=\gm\X-\al'\Y-\be'\Z\sin\al,\qquad
g=\al'\X+\gm\Y-\be'\Z\cos\al,\eqno(3.46)$$
$$h=-(\be'\X\sin\al+\be'\Y\cos\al+2\gm\Z)\eqno(3.47)$$ modulo the
transformations of type in (1.8) and (1.9) for some function $\gm$
of $t$. With the above data, we have:
\begin{eqnarray*}\hspace{1.1cm}R_1&=&(\gm'+\gm^2-(\al')^2+
3(\be')^2\sin^2\al)\X +12\nu \al'\Y\X^{-2}-24\nu^2\X^{-3}
\\ & &+(3(\be')^2\sin\al\;\cos \al-{\al'}'-2\al'\gm)\Y
+(4\be'\gm-{\be'}')\Z\sin\al, \hspace{2cm}(3.48)\end{eqnarray*}
\begin{eqnarray*}\hspace{1.1cm}R_2&=&(\gm'+\gm^2-(\al')^2+
3(\be')^2\cos^2\al)\Y-12\nu \al'\X^{-1}\\&
&+({\al'}'+2\al'\gm+3(\be')^2\sin\al\;\cos \al)\X
+(4\be'\gm-{\be'}')\Z\cos\al,\hspace{1.9cm}(3.49)\end{eqnarray*}
$$R_3=(4\gm^2-2\gm'-(\be')^2)\Z+(4\be'\gm-{\be'}')
(\X\sin\al+\Y\cos\al).\eqno(3.50)$$

By (3.48)-(3.50), (3.14) is now equivalent to
$$-{\al'}'-2\al'\gm
={\al'}'+2\al'\gm\lra\gm=-\frac{{\al'}'}{2\al'}.\eqno(3.51)$$ Thus
$$\U=-\frac{{\al'}'}{2\al'}\X-\al'\Y-\be'\Z\sin\al-6\nu\X^{-1},\qquad
\V=\al'\X-\frac{{\al'}'}{2\al'}\Y-\be'\Z\cos\al-6\nu\Y\X^{-2},\eqno(3.52)$$
$$\W=\frac{{\al'}'}{\al'}\Z-\be'\X\sin\al-\be'\Y\cos\al\eqno(3.53)$$
by (3.42), (3.46), (3.47) and (3.51). Moreover, (3.12) and
(3.48)-(3.50) imply
\begin{eqnarray*}p&=&\rho\{
\frac{(2\al'{{\al'}'}'+4(\al')^4-3({\al'}')^2)(\X^2+\Y^2)}{8(\al')^2}-
\frac{3(\be')^2(\X^2\sin^2\al+\Y^2\cos^2\al)}{2}
\\ & &+12\nu(\al'\Y\X^{-1}-\nu\X^{-2})
+({\be'}'-4\be'\gm)\Z(\X\sin\al+\Y\cos\al)\\ &
&-3(\be')^2\X\Y\sin\al\;\cos \al+
\frac{(\al'{{\al'}'}'+(\al')^2(\be')^2-2({\al'}')^2)\Z^2}{2(\al')^2}
\}.\hspace{3cm}(3.54)\end{eqnarray*} By (3.3) and (3.5), we have
the following theorem:\psp

 {\bf Theorem 3.3}. {\it Let
$\al$ and $\be$ be functions of $t$ with $\al'\neq 0$. In terms of
the notations $\X,\Y,\Z$ given in (3.1) and (3.5), we have the
following solution of  Navier-Stokes equations (1.1)-(1.4):
$$u=\left(\frac{{\al'}'}{2\al'}+6\nu\Y\X^{-2}\right)(\Y\sin\al-\X\cos\al)
-\al'(\X\sin\al+\Y\cos\al),\eqno(3.55)$$
\begin{eqnarray*}\hspace{0.6cm}v&=&-\left(\frac{{\al'}'}
{2\al'}+6\nu\Y\X^{-2}\right)(\X\sin\al+\Y\cos\al)\cos\be
+\al'(\X\cos\al-\Y\sin\al)\cos\be\\ &&-\be'\Z\cos\be
+\left(\be'\X\sin\al+\be'\Y\cos\al-\frac{{\al'}'}{\al'}\Z\right)\sin\be,
\hspace{3.7cm}(3.56)\end{eqnarray*}
\begin{eqnarray*}\hspace{0.6cm}w&=&-\left(\frac{{\al'}'}
{2\al'}+6\nu\Y\X^{-2}\right)(\X\sin\al+\Y\cos\al)\sin\be
+\al'(\X\cos\al-\Y\sin\al)\sin\be\\ &&-\be'\Z\sin\be
+\left(\frac{{\al'}'}{\al'}\Z-\be'\X\sin\al-\be'\Y\cos\al\right)\cos\be
\hspace{3.8cm}(3.57)\end{eqnarray*}
 and $p$ is given in (3.54). The above solution blows up at any
 point on the following rotating plane with a line deleted:
 \begin{eqnarray*}\hspace{2cm}& &\{(x,y,z)\in\mbb{R}^3\mid x\cos\al+y\sin\al\;\cos\be
 +z\sin\al\;\sin\be=0,\\ & &-x\sin\al+y\cos\al\;\cos\be+z\cos\al\;\sin\be\neq
 0\}.\hspace{4.2cm}(3.58)\end{eqnarray*}}
\pse

We remark that the above solution may be applied to study
turbulence.

\section{Moving-Frame Approach II}

Motivated from the solution $\psi$ in (2.27) of the equation
(2.24), we will solve Navier-Stokes equations by ansatzes with
given irrational functions under the moving frames in (3.5).

First we rewrite (3.9)-(3.11):
\begin{eqnarray*}& &R_1=\U_t+(\al'\Y+\be'\Z\sin\al+\U)\U_\X
+(\V-\al'\X+\be'\Z\cos\al)\U_\Y \\
&&+(\W-\be'(\X\sin\al+\Y\cos\al))\U_\Z-\al'\V-\be'\W\sin\al-\nu\Dlt(\U),
\hspace{3.2cm}(4.1)\end{eqnarray*}
\begin{eqnarray*}& &R_2=\V_t+(\al'\Y+\be'\Z\sin\al+\U)\V_\X
+(\V-\al'\X+\be'\Z\cos\al)\V_\Y \\
&&+(\W-\be'(\X\sin\al+\Y\cos\al))\V_\Z+\al'\U-\be'\W\cos\al-\nu\Dlt(\V),
\hspace{3.2cm}(4.2)\end{eqnarray*}
\begin{eqnarray*}& &R_3=\W_t+(\al'\Y+\be'\Z\sin\al+\U)\W_\X
+(\V-\al'\X+\be'\Z\cos\al)\W_\Y \\
&&+(\W-\be'(\X\sin\al+\Y\cos\al))\W_\Z+
\be'(\U\sin\al+\V\cos\al)-\nu\Dlt(\W).
\hspace{2.3cm}(4.3)\end{eqnarray*} Let $\al_1,\be_1,\gm$ be
functions of $t$, and let $a,b$ be  real numbers. Set
$$\xi_0=e^{\al_1\Y+\be_1\Z}-ae^{-\al_1\Y-\be_1\Z},\qquad\zeta_0=
e^{\al_1\Y+\be_1\Z}+ae^{-\al_1\Y-\be_1\Z},\eqno(4.4)$$
$$\xi_1=\sin(\al_1\Y+\be_1\Z),\qquad\zeta_1=\cos(\al_1\Y+\be_1\Z),
\eqno(4.5)$$
$$\phi_0=e^{\gm\X}-be^{-\gm\X},\qquad\zeta_0=
e^{\gm\X}+be^{-\gm \X},\eqno(4.6)$$ $$\phi_1=\sin(\gm\X),\qquad
\psi_1=\cos(\gm\X),\qquad\Dlt_1=\ptl_\Y^2+\ptl_\Z^2.\eqno(4.7)$$

Suppose that $f$ and $h$ are functions in $t,\Y,\Z$. According to
(3.46) and (3.47), we assume
$$\U=-\al'\Y-\be'\Z\sin\al-(f_\Y+h_\Z)\X
-(\al_1\sgm+\be_1\tau)\zeta_r\phi_s,\eqno(4.8)$$
$$\V=\al'\X-\be'\Z\cos\al+f+\sgm\gm\xi_r\psi_s,\qquad
\W=\be'(\X\sin\al+\Y\cos\al)+h+\tau\gm\xi_r\psi_s.\eqno(4.9)$$ By
(4.1)-(4.3), we have
\begin{eqnarray*}&
&R_1= -(\al_1\sgm+\be_1\tau)'\zeta_r\phi_s-
(\al_1\sgm+\be_1\tau)[(-1)^r(\al_1'\Y+\be_1'\Z)\xi_r\phi_s+\gm'\X\zeta_r\psi_s]
\\ & &-(f_{\Y t}+h_{\Z t})\X+((f_\Y+h_\Z)\X+(\al_1\sgm+\be_1\tau)\zeta_r\phi_s)(
f_\Y+h_\Z+\gm(\al_1\sgm+\be_1\tau)\zeta_r\psi_s)
\\ & &-\al'(f+\al'\X-\be'\Z\cos\al+\gm\sgm\xi_r\psi_s)
-\be'(\be'(\X\sin\al+\Y\cos\al)+h+\gm\tau\xi_r\psi_s)\sin\al
\\ &&-(f+\gm\sgm\xi_r\psi_s)(\al'+(f_{\Y\Y}+h_{\Y\Z})\X+(-1)^r\al_1(\al_1\sgm+\be_1\tau)\xi_r\phi_s)
-(h+\gm\tau\xi_r\psi_s)\\ & &\times
(\be'\sin\al+(f_{\Y\Z}+h_{\Z\Z})\X+(-1)^r\be_1(\al_1\sgm+\be_1\tau)\xi_r\phi_s)
+\nu\{\Dlt_1(f_\Y+h_\Z)\X
\\ & &+(\al_1\sgm+\be_1\tau)[(-1)^r(\al_1^2+\be_1^2)+(-1)^s\gm^2]
\zeta_r\phi_s\}-{\al'}'\Y-(\be'\sin\al)'\Z\\&
&=\{(\gm(f_\Y+h_\Z)-\gm')\X\zeta_r\psi_s-
(-1)^r(\al_1'\Y+\be_1'\Z+\al_1f+\be_1h
)\xi_r\phi_s\}(\al_1\sgm+\be_1\tau)
\\ & &+\{(\al_1\sgm+\be_1\tau)[(-1)^s\nu\gm^2
+(-1)^r\nu(\al_1^2+\be_1^2)+f_\Y+h_\Z]-(\al_1\sgm+\tau\be_1)'\}\zeta_r\phi_s
\\ &&-\gm\{2(\sgm\al'+\tau\be'\sin\al)
+[\sgm(f_{\Y\Y}+h_{\Y\Z})+\tau(f_{\Y\Z}+h_{\Z\Z})]\X\} \xi_r\psi_s
-(f_{\Y t}+h_{\Z t})\X\\
&&+(f_\Y+h_\Z)^2\X-f(\al'+(f_{\Y\Y}+h_{\Y\Z})\X)
-h(\be'\sin\al+(f_{\Y\Z}+h_{\Z\Z})\X)
\\ & &-\al'(f+\al'\X-\be'\Z\cos\al)
-\be'(\be'(\X\sin\al+\Y\cos\al)+h)\sin\al -{\al'}'\Y\\ &
&-(\be'\sin\al)'\Z+\nu\Dlt_1(f_\Y+h_\Z)\X+\gm(\al_1\sgm+\be_1\tau)^2(\dlt_{r,1}+4a\dlt_{r,0})\phi_s\psi_s
, \hspace{2cm}(4.10)\end{eqnarray*}
\begin{eqnarray*}&
&R_2={\al'}'\X-(\be'\cos\al)'\Z+
f_t+(\gm\sgm)'\xi_r\psi_s+\gm\sgm((\al_1'\Y+\be_1'\Z)\zeta_r\psi_s
+(-1)^s\gm'\X\xi_r\phi_s)\\ &
&-\al'[\al'\Y+\be'\Z\sin\al+(f_\Y+h_\Z)\X+(\al_1\sgm+\be_1\tau)\zeta_r\phi_s]
-\be'[\be'(\X\sin\al+\Y\cos\al)\\ & &+
h+\gm\tau\xi_r\psi_s]\cos\al
-[(f_\Y+h_\Z)\X+(\al_1\sgm+\be_1\tau)\zeta_r\phi_s]
(\al'+(-1)^s\gm^2\sgm\xi_r\phi_s)
\\ & &+(f+\gm\sgm\xi_r\psi_s)(f_\Y+
\al_1\gm\sgm\zeta_r\psi_s)+(h+\gm\tau\xi_r\psi_s)(f_\Z-\be'\cos\al
+\be_1\gm\sgm\zeta_r\psi_s)\\& &
-\nu[\Dlt_1(f)+\gm\sgm((-1)^r(\al_1^2+\be_1^2)+(-1)^s\gm^2)]\xi_r\psi_s
\\ &&={\al'}'\X+f_t+\gm\sgm(\al_1'\Y+\be_1'\Z+\al_1f+\be_1h) \zeta_r\psi_s
+(-1)^s\gm\sgm(\gm'-\gm(f_\Y+h_\Z))\X \xi_r\phi_s \\ &
&+\{\gm[\sgm f_\Y+\tau f_\Z
-\nu\sgm[(-1)^r(\al_1^2+\be_1^2)+(-1)^s\gm^2]
-2\tau\be'\cos\al]+(\gm\sgm)'\}\xi_r\psi_s
\\ & &-2\al'(\al_1\sgm+\be_1\tau)\zeta_r\phi_s-(\be'\cos\al)'\Z
-\al'[\al'\Y+\be'\Z\sin\al+2(f_\Y+h_\Z)\X]
 \\ &
 &-\be'[\be'(\X\sin\al+\Y\cos\al)+h]\cos\al+ff_\Y+h(f_\Z-\be'\cos\al)
-\nu\Dlt_1(f)\\ & &
+\gm^2\sgm(\al_1\sgm+\be_1\tau)(4b\dlt_{s,0}+\dlt_{s,1})\xi_r\zeta_r
,  \hspace{7.9cm}(4.11)\end{eqnarray*}
\begin{eqnarray*}& &R_3=(\be'\sin\al)'\X+(\be'\cos\al)'\Y+h_t
+(\tau\gm)'\xi_r\psi_s+ \tau\gm[(\al_1'\Y+\be_1'\Z)
\zeta_r\psi_s\\ & &+(-1)^s\gm'\X\xi_r\phi_s] -[(f_\Y+h_\Z)\X+
(\al_1\sgm+\be_1\tau)\zeta_r\phi_s](\be'\sin\al+(-1)^s
\tau\gm^2\xi_r\phi_s)\\ & &+(f+\sgm\gm\xi_r\psi_s)
(\be'\cos\al+h_\Y+\al_1\tau\gm\zeta_r\psi_s)+(h+\tau\gm\xi_r\psi_s)
(h_\Z+\be_1\tau\gm\zeta_r\psi_s)\hspace{4cm}\end{eqnarray*}\begin{eqnarray*}
& &-\be'(\al'\Y+\be'\Z\sin\al+(f_\Y+h_\Z)\X+(\al_1\sgm+\be_1\tau)
\zeta_r\phi_s)\sin\al +\be'(\al'\X-\be'\Z\cos\al\\ &
&+f+\sgm\gm\xi_r\psi_s)\cos\al
-\nu[\Dlt_1(h)+\gm\tau((-1)^r(\al_1^2+\be_1^2)
+(-1)^s\gm^2)]\xi_r\psi_s
\\ &&=\gm\tau(\al_1'\Y+\be_1'\Z+\al_1f+\be_1h)\zeta_r\psi_s
+\{(\tau\gm)'-\nu\gm\tau[(-1)^r(\al_1^2+\be_1^2) +(-1)^s\gm^2]\\
&&+\gm(2\be'\sgm\cos\al+\sgm h_\Y+\tau h_\Z) \}\xi_r\psi_s
-2\be'(\al_1\sgm+\be_1\tau)\zeta_r\phi_s\sin\al+
(-1)^s\gm\tau(\gm'\\ & &-\gm(f_\Y+h_\Z))\X\xi_r\phi_s
+(\be'\sin\al)'\X+(\be'\cos\al)'\Y+h_t-\be'(f_\Y+h_\Z)\X\sin\al
\\ &&+f(\be'\cos\al+h_\Y)+hh_\Z-
\be'(\al'\Y+\be'\Z\sin\al+(f_\Y+h_\Z)\X)\sin\al+\be'(\al'\X\\ &
&-\be'\Z\cos\al+f)\cos\al -\nu\Dlt_1(h)
+\gm^2\tau(\al_1\sgm+\be_1\tau)(4b\dlt_{s,0}+\dlt_{s,1})\xi_r\zeta_r.
\hspace{2.1cm}(4.12)\end{eqnarray*}

By the coefficients of $\xi_r\psi_s$ in the equation
$\ptl_\Y(R_1)=\ptl_\X(R_2)$, we have
$$\gm^2\sgm=(-1)^{r+s+1}\al_1(\al_1\sgm+\be_1\tau),\qquad
[\sgm(f_{\Y\Y}+h_{\Y\Z})+\tau(f_{\Y\Z}+h_{\Z\Z})]_\Y=0.\eqno(4.13)$$
Moreover, the coefficients of $\zeta_r\phi_s$ in the equation
$\ptl_\Y(R_1)=\ptl_\X(R_2)$ suggest
$$(f_\Y+h_\Z)_\Y=0,\eqno(4.14)$$
which implies the second equation in (4.13). According the
coefficients of $\xi_r\phi_s$ in the equation
$\ptl_\Y(R_1)=\ptl_\X(R_2)$, we get
$$\sgm\be_1h_\Y =\al_1(\tau f_\Z -2\tau\be'\cos\al).\eqno(4.15)$$
Furthermore, the coefficients of $\zeta_r\psi_s$ in the equation
$\ptl_\Y(R_1)=\ptl_\X(R_2)$ yield
$$\al_1\be'\sin\al=\al'\be_1.\eqno(4.16)$$
Symmetrically, we have (4.16),
$$\gm^2\tau=(-1)^{r+s+1}\be_1(\al_1\sgm+\be_1\tau),\qquad
(f_\Y+h_\Z)_\Z=0\eqno(4.17)$$ and
$$f_\Z
= h_\Y+2\be'\cos\al.\eqno(4.18)$$ By the first equation in (4.13)
and (4.17), we have
$$\sgm\be_1=\tau\al_1.\eqno(4.19)$$
Then (4.15) is implied by (4.18) and (4.19). Note that the
equations of  the coefficients
$\xi_r\psi_s,\;\zeta_r\psi_s,\;\xi_r\phi_s$ and $\zeta_r\phi_s$ in
$\ptl_\Z(R_2)=\ptl_\Y(R_3)$ are implied by (4.16), (4.18) and
(4.19).

According to (4.14) and the second equation in (4.17),
$$f_\Y+h_\Z=\gm_1,\eqno(4.20)$$
a function of $t$. Under the conditions in (4.16), the first
equation in (4.17), and (4.18)-(4.20), $\ptl_\Y(R_1)=\ptl_\X(R_2)$
becomes
$$ \al'h_\Z-\be'h_\Y\sin\al
={\al'}',\eqno(4.21)$$ $\ptl_\Z(R_1)=\ptl_\X(R_3)$ is equivalent
to
$$\be'h_\Z\sin\al+\al'
h_y=\be'\gm_1\sin\al-(\be'\sin\al)'-2\al'\be'\cos\al \eqno(4.22)$$
and $\ptl_\Z(R_2)=\ptl_\Y(R_3)$ says
$$(ff_\Y+hf_\Z)_\Z
=(fh_\Y+hh_\Z)_\Y +2\be'\gm_1\cos\al.\eqno(4.23)$$ By (4.18) and
(4.20)-(4.22), we assume $f_\Y,\;f_\Z,\;h_\Y$ and $h_\Z$ are
functions of $t$. Then (4.23) can be written as
$$(f_\Y+h_\Z)f_\Z=(f_\Y+h_\Z)h_\Y+2\be'\gm_1\cos\al,\eqno(4.24)$$
which is implied by (4.18) and (4.20). Solving (4.21) and (4.22),
we get
$$h_\Y=\frac{\al'\be'\gm_1\sin\al-(\al'\be'\sin\al)'
-2(\al')^2\be'\cos\al}{(\al')^2+(\be')^2\sin^2\al},\eqno(4.25)$$
$$h_\Z= \frac{\al'{\al'}'+(\be')^2\gm_1\sin^2\al-
(\be'\sin\al)(\be'\sin\al)'-\al'(\be')^2\sin
2\al}{(\al')^2+(\be')^2\sin^2\al}.\eqno(4.26)$$ Moreover,
$$f_\Y=\frac{\gm_1(\al')^2-\al'{\al'}'+
(\be'\sin\al)(\be'\sin\al)'+\al'(\be')^2\sin
2\al}{(\al')^2+(\be')^2\sin^2\al}\eqno(4.27)$$ by (4.20) and
(4.26), and
$$f_\Z=\frac{\al'\be'\gm_1\sin\al-(\al'\be'\sin\al)'
+2(\be')^2\sin^2\al\;\cos\al}{(\al')^2+(\be')^2\sin^2\al}
\eqno(4.28)$$ by (4.18) and (4.25). With above data, we take
$$f=f_\Y\Y+f_\Z\Z,\qquad h=h_\Y\Y+h_\Z\Z\eqno(4.29)$$
by the transformations of the type in (1.8) and (1.9).
 Furthermore, (4.17), (4.19)
and the first equation in (4.17) yield $r+s+1\in 2\mbb{Z}$,
$$\al_1=\vf\al',\qquad \gm=\pm\vf\sqrt{(\al')^2+(\be')^2\sin^2\al},
\eqno(4.30)$$
$$\be_1=\vf\be'\sin\al,\qquad\sgm=\mu\al',\qquad
\tau=\mu\be'\sin\al.\eqno(4.31)$$ In particular,
$\al,\be,\gm_1,\vf$ and $\mu$ are arbitrary functions of $t$.
 According (4.9)-(4.11), the pressure
\begin{eqnarray*}&
&p=\rho \{\gm\mu\vf^{-1}[(\gm'-\gm\gm_1)\X\zeta_r\phi_s
-((\vf\al')'\Y+(\vf\be'\sin\al)'\Z+\vf(\al'f+\be'h\sin\al) )\xi_r\psi_s] \\
& & +(-1)^s\vf^{-1}[(\gm\mu)'-\gm\mu\vf'\vf^{-1}]\zeta_r\psi_s
+2\mu((\al')^2+(\be')^2\sin^2\al)\xi_r\phi_s+2\al'f\X\\
& & +\frac{(\al')^2+(\be')^2\sin^2\al+\gm_1'-\gm_1^2}{2}\X^2
 +2\be'h\X\sin\al +[(\be'\sin\al)'-\al'\be'\cos\al]\X\Z
\\ & & +\left(\frac{(\be')^2}{2}\sin2\al+{\al'}'\right)\X\Y-
\frac{1}{2}\gm^4\mu^2\vf^{-2}[(\dlt_{r,1}+4a\dlt_{r,0})\phi_s^2
+(4b\dlt_{s,0}+\dlt_{s,1})\xi_r^2]
\\ & &+[(\be'\cos\al)'+\al'\be'\sin\al-f_{\Z t}-f_\Y f_\Z-h_\Y h_\Z]\Y\Z+\frac{(\be')^2-h_{\Z t}-f_\Z^2-h_\Z^2}{2}\Z^2
\\ & &+ \frac{(\al')^2+(\be')^2\cos\al-f_{\Y t}-f_\Y^2-h_\Y^2}{2}\Y^2
\}\hspace{6.9cm}(4.32)
\end{eqnarray*}
modulo the transformation in (1.10). By (3.3) and (3.5), we have
the following theorem: \psp

{\bf Theorem 4.1}. {\it Let $\al,\be,\gm_1,\vf$ and $\mu$ be
arbitrary functions of $t$ such that $\vf\neq 0$ and
$(\al')^2+(\be')^2\sin^2\al\neq 0$. Take any real constants $a$
and $b$. The notations $\X,\Y$ and $\Z$ are defined in (3.5) via
(3.1),  and the notations $\xi_r,\zeta_r,\phi_r$ and $\psi_r$ are
defined in (4.4)-(4.7) with $\al_1,\be_1$ and $\gm$ given in
(4.30) and (4.31). Moreover, $f_\Y,f_\Z,h_\Y,h_\Z$ and $f, h$ are
given in (4.25)-(4.29). Assume $(r,s)\in\{(0,1),(1,0)\}$. We have
the following solution of the Navier-Stokes equations (1.1)-(1.4):
\begin{eqnarray*}\hspace{1.7cm} u&=&-\al'(\X\sin\al+\Y\cos\al)
-(f+\mu\al'\gm\xi_r\psi_s)\sin\al\\ & & -(\gm_1\X+ \vf
\mu((\al')^2+(\be')^2\sin^2\al)\zeta_r\phi_s)\cos\al,\hspace{4cm}(4.33)
\end{eqnarray*}
\begin{eqnarray*}\hspace{1.2cm}v&=&(f\cos\al-\be'\Z)\cos\be
-(\al'\sin\al\;\cos\be+\be'\cos\al\;\sin\be)\Y
\\ & &-(\gm_1\X+ \vf
\mu((\al')^2+(\be')^2\sin^2\al)\zeta_r\phi_s)\sin\al\;\cos\be
-h\sin\be\\ & &+(\al'\cos\al\;\cos\be-
\be'\sin\al\;\sin\be)(\X+\gm\mu\xi_r\psi_s),\hspace{4cm}(4.34)\end{eqnarray*}
\begin{eqnarray*}\hspace{1.2cm}w&=&(\be'\cos\al\;\cos\be-
\al'\sin\al\;\sin\be)\Y +(f\cos\al-\be'\Z)\sin\be\\& & -(\gm_1\X+
\vf \mu((\al')^2+(\be')^2\sin^2\al)\zeta_r\phi_s)\sin\al\;\sin\be
+h\cos\be\\&&+(\al'\cos\al\;\sin\be+\be'\sin\al\;\cos\be)(\X+\gm\mu\xi_r\psi_s)
\hspace{4cm}(4.35)\end{eqnarray*} and $p$ is given in (4.32). The
above solution is globally analytic in $x,y,z$.}\psp

Let $\gm_1,\gm_2$ be functions of $t$ and let $a,b,c$ be real
numbers. Denote
$$\phi_0=
e^{\gm_1\Y+\gm_2\Z}-ae^{-\gm_1\Y-\gm_2\Z},
\qquad\phi_1=\sin(\gm_1\Y+\gm_2\Z),\eqno(4.36)$$
$$\psi_0=
e^{\gm_1\Y+\gm_2\Z}+ae^{-\gm_1\Y-\gm_2\Z},
\qquad\psi_1=\cos(\gm_1\Y+\gm_2\Z),\eqno(4.37)$$
$$\xi_0=
be^{\gm_1\Y+\gm_2\Z}-ce^{-\gm_1\Y-\gm_2\Z},
\qquad\xi_1=c\sin(\gm_1\Y+\gm_2\Z+b),\eqno(4.38)$$
$$\zeta_0=
be^{\gm_1\Y+\gm_2\Z}+ce^{-\gm_1\Y-\gm_2\Z},
\qquad\zeta_1=c\cos(\gm_1\Y+\gm_2\Z+b).\eqno(4.39)$$ Suppose that
$\sgm,\tau$ are functions of $t$ and $f,k,h$ are functions in
$t,\X,\Y,\Z$ such that $h$ and $g$ are linear in $\X,Y,\Z$ and
$$f_\X+k_\Y+h_\Z=0.\eqno(4.40)$$
Motivated from the above solution, we consider the solution of the
form:
$$\U=-\al'\Y-\be'\Z\sin\al+f-(\gm_1^2+\gm_2^2)
(\tau\zeta_r\X+\sgm\psi_r\X^2),\eqno(4.41)$$
$$\V=\al'\X-\be'\Z\cos\al+k+\gm_1(\tau\xi_r+2\sgm\phi_r\X),\eqno(4.42)$$
$$\W=\be'(\X\sin\al+\Y\cos\al)+h+\gm_2(\tau\xi_r+2\sgm\phi_r\X)
.\eqno(4.43)$$

For convenience of computation, we denote
$$\gm=\gm_1^2+\gm_2^2,\qquad f^\ast=f-f_x\X
\qquad\Dlt_1=\ptl_\Y^2+\ptl_\Z^2.\eqno(4.44)$$ Now (4.1) becomes
\begin{eqnarray*}&
&R_1=-{\al'}'\Y-(\be'\sin\al)'\Z+f_t-(-1)^r\gm(\gm_1'\Y+\gm_2'\Z)(\tau\xi_r\X+\sgm\phi_r\X^2)\\
& & +((-1)^r\nu \gm^2\tau-(\gm\tau)')\zeta_r\X +
(f-\gm(\tau\zeta_r\X+\sgm\psi_r\X^2))(f_\X-\gm(\tau\zeta_r+2\sgm\psi_r\X))
\\ & &+(k+\gm_1(\tau\xi_r+2\sgm\phi_r\X)) [f_\Y-2\al'-(-1)^r\gm\gm_1
(\tau\xi_r\X+\sgm\phi_r\X^2)]-\nu\Dlt_1(f)\\ & &
+(h+\gm_2(\tau\xi_r+2\sgm\phi_r\X))[f_\Z-2\be'\sin\al-(-1)^r\gm\gm_2
(\tau\xi_r\X+\sgm\phi_r\X^2)]+2\nu\gm\sgm\psi_r\\&&-\al'(\al'\X-\be'\Z\cos\al)
-(\be')^2(\X\sin\al+\Y\cos\al)\sin\al+((-1)^r\nu
\gm^2\sgm-(\gm\sgm)')\psi_r\X^2
\\&&=-((\al')^2+(\be')^2\sin^2\al)\X-({\al'}'+2^{-1}(\be')^2\sin2\al)\Y
+(\al'\be'\cos\al-(\be'\sin\al)')\Z\\ &
&+\gm^2[\tau^2(4b\dlt_{0,r}+c\dlt_{1,r})c\X+3\sgm\tau(2\dlt_{0,r}(ab+c)+
\dlt_{1,r}c\cos b)\X^2+2\sgm^2 (4a\dlt_{0,r}+\dlt_{1,r})\X^3]
\\ & &-(-1)^r\gm(\gm_1'\Y+\gm_2'\Z+k\gm_1+h\gm_2)
(\tau\xi_r\X+\sgm\phi_r\X^2)+ff_\X+kf_\Y+hf_\Z\\ & &+((-1)^r\nu
\gm^2\sgm-(\gm\sgm)'-3\gm\sgm
f_\X)\psi_r\X^2+\nu(2\sgm\psi_r-\Dlt_1(f))-\gm\tau f^\ast\zeta_r\\
& &-[((\gm\tau)'+2\gm\tau f_\X-(-1)^r\nu
\gm^2\tau)\zeta_r+2\gm\sgm
f^\ast\psi_r]\X+f_t\\
&& +(\gm_1 (f_\Y-2\al') +\gm_2(f_\Z-2\be'\sin\al))
(\tau\xi_r+2\sgm\phi_r\X).\hspace{4.8cm}(4.45)
\end{eqnarray*}
To solve (3.14), we assume
$$\gm_1'\Y+\gm_2'\Z+k\gm_1+h\gm_2=0\eqno(4.46)$$
and
$$
(-1)^r\nu\gm^2\sgm-(\gm\sgm)'-3\gm\sgm f_\X=0,\eqno(4.47)$$
 Moreover, (4.2) and
(4.3) become
\begin{eqnarray*}& &R_2={\al'}'\X-(\be'\cos\al)'\Z+
((\gm_1\tau)'-(-1)^r\nu\gm\gm_1\tau)\xi_r+2((\gm_1\sgm)'-
(-1)^r\nu\gm\gm_1\sgm)\phi_r\X\\ &
&+k_t+(\gm_1'\Y+\gm_2'\Z)\gm_1(\tau\zeta_r+2\sgm\psi_r\X)
+(f-\gm(\tau\zeta_r\X+\sgm\psi_r\X^2))(2\al'+k_\X+2\gm_1\sgm\phi_r)
\\ & &+(k+\gm_1(\tau\xi_r+2\sgm\phi_r\X))(k_\Y+\gm_1^2(\tau\zeta_r
+2\sgm\psi_r\X))-(\be')^2(\X\sin\al+\Y\cos\al)\cos\al\\ & &
-\al'(\al'\Y+\be'\Z\sin\al)+(h+\gm_2(\tau\xi_r+2\sgm\phi_r\X))(k_\Z
-2\be'\cos\al+\gm_1\gm_2
(\tau\zeta_r+2\sgm\psi_r\X))\\
&&=({\al'}'-2^{-1}(\be')^2\sin2\al+f_\X(2\al'+k_\X))\X
-((\al')^2+(\be')^2\cos^2\al)\Y+k_t+kk_\Y\\ & & +
[\tau(\gm_1k_\Y+\gm_2(k_\Z-2\be'\cos\al))+(\gm_1\tau)'-(-1)^r\nu\gm\gm_1\tau]\xi_r
-((\be'\cos\al)'+\al'\be'\sin\al)\Z\\&&+\gm\sgm(2\gm_1\sgm\phi_r-2\al'
-k_\X)\psi_r\X^2+f^\ast(2\al'+k_\X+2\gm_1\sgm\phi_r)+h(k_\Z-2\be'\cos\al)
\\ &&+\gm\gm_1\tau^2\xi_r\zeta_r
+ \{2\gm\gm_1\sgm\tau\xi_r\psi_r +2[(\gm_1\sgm)'-\sgm(\gm_1(h_\Z+
(-1)^r\nu\gm)\\ & &+\gm_2(2\be'\cos\al-k_\Z))]\phi_r
-\gm\tau(2\al'+k_\X)\zeta_r\}\X\hspace{6.2cm}(4.48)\end{eqnarray*}
\begin{eqnarray*}& &R_3=(\be'\sin\al)'\X+(\be'\cos\al)'\Y+
 (\gm_2\tau)'\xi_r+2(\gm_2\sgm)'\phi_r\X-(-1)^r\nu\gm_2^3(\tau\xi_r+\sgm\phi_r\X)\\ &
&+(\gm_1'\Y+\gm_2'\Z)\gm_2(\tau\zeta_r+2\sgm\psi_r\X)
+(f-\gm(\tau\zeta_r\X+\sgm\psi_r\X^2))(2\be'\sin\al+h_\X+2\gm_2\sgm\phi_r)
\\ & &+(k+\gm_1(\tau\xi_r+2\sgm\phi_r\X))(2\be'\cos\al+
h_\Y+\gm_1\gm_2(\tau\zeta_r +2\sgm\psi_r\X)-(\be')^2\Z+h_t\\ &
&+\al'\be'(\X\cos\al-\Y\sin\al)
+(h+\gm_2(\tau\xi_r+2\sgm\phi_r\X))(h_\Z
+\gm_2^2(\tau\zeta_r+2\sgm\psi_r\X))\\ &
&=[(\be'\sin\al)'+\al'\be'\cos\al+f_\X(2\be'\sin\al+h_\X)]\X+
[(\be'\cos\al)'-\al'\be'\sin\al]\Y
\\ & & +[(\gm_2\tau)'+(\gm_1(2\be'\cos\al+
h_\Y)+\gm_2h_\Z-(-1)^r\nu\gm\gm_2)\tau]\xi_r+\{
2\gm\gm_2\tau\sgm\xi_r\psi_r +2[(\gm_2\sgm)'\\ &
&-\gm_2\sgm(k_\Y+(-1)^r\nu\gm)+\gm_1\sgm
(2\be'\cos\al+h_\Y)]\phi_r-\gm\tau(2\be'\sin\al+ h_\X)\zeta_r\}\X\\
& &+f^\ast(2\be'\sin\al+h_\X+2\gm_2\sgm\phi_r) +k(2\be'\cos\al+
h_\Y)+h_t+hh_\Z+\gm\gm_2\tau^2\xi_r\zeta_r\\ &
&+\gm\sgm\psi_r(2\gm_2\sgm\phi_r-2\be'\sin\al-h_\X)\X^2-(\be')^2\Z.
\hspace{6cm}(4.49)
\end{eqnarray*}

By the coefficients of $\X^2$ in $\ptl_\Z(R_2)=\ptl_\Y(R_3)$, we
have:
$$\gm_2(2\al'+k_\X)=\gm_1(2\be'\sin\al+h_\X).\eqno(4.50)$$
According to (4.46),
$$k_\X\gm_1+h_\X\gm_2=0,\;\;\gm_1'+\gm_1k_\Y+\gm_2h_\Y=0,\;\;
\gm_2'+\gm_1k_\Z+\gm_2h_\Z=0.\eqno(4.51)$$ Solving (4.50) and the
first equation in (4.51), we obtain
$$k_\X=2\gm^{-1}\gm_2(\be'\gm_1\sin\al-\al'\gm_2),\qquad h_\X=
-2\gm^{-1}\gm_1(\be'\gm_1\sin\al-\al'\gm_2).\eqno(4.52)$$
Moreover, the coefficients of $\X$ in $\ptl_\Z(R_2)=\ptl_\Y(R_3)$
give
$$\gm_1'\gm_2-\gm_1\gm_2'+\gm_1\gm_2(k_\Y-h_\Z
)+\gm_2^2k_\Z-\gm_1^2h_\Y-2\gm\be'\cos\al=0. \eqno(4.53)$$ By
(4.51), the above equation can be rewritten as
$$k_\Z-h_\Y=2\be'\cos\al.
\eqno(4.54)$$ Furthermore,  (4.50) and the coefficients of $\X^0$
in $\ptl_\Z(R_2)=\ptl_\Y(R_3)$ show that $f$ is a function of $t$
and $\gm_1\Y+\gm_2\Z$.
 According to the coefficients of $\X$ in
$\ptl_\Y(R_1)=\ptl_\X(R_2)$ and $\ptl_\Z(R_1)=\ptl_\X(R_3)$, we
take
$$f^\ast=\vf\vt_r+\sgm\td\varpi\phi_r+\al_1
,\eqno(4.55)$$ where $\vf$ and $\al_1$ are functions of $t$, and
$$\td\varpi=\gm_1\Y+\gm_2\Z,\qquad
\vt_0=b_1e^{\td\varpi}-c_1e^{-\td\varpi},\qquad
\vt_1=c_1\sin(\td\varpi+b_1)\eqno(4.56)$$ for $b_1,c_1\in\mbb{R}$.

Now the coefficients of $\X$ in $\ptl_\Y(R_1)=\ptl_\X(R_2)$ and
$\ptl_\X(R_1)=\ptl_\X(R_3)$ give
$${\cal T}[2\al_1\gm\sgm\psi_r+((\gm\tau)'+2\gm\tau f_\X-(-1)^r\nu
\gm^2\tau)\zeta_r]=0,\qquad{\cal T}=\ptl_\Y,\ptl_\Z.\eqno(4.57)$$
 Moreover, the coefficients of $\X^0$ in
$\ptl_\Y(R_1)=\ptl_\X(R_2)$ and $\ptl_\X(R_1)=\ptl_\X(R_3)$ yield
\begin{eqnarray*}&&[(f_\X-(-1)^r\gm\nu)\vf+\vf')\vt_r
+((f_\X-(-1)^r\nu\gm)\sgm+\sgm')\td\varpi\phi_r
-\al_1\gm\tau\zeta_r]_\Y\\
 &=&2[(\gm_1\sgm)'-\sgm(\gm_1(h_\Z+
(-1)^r\nu\gm)-\gm_2h_\Y)]\phi_r\\
& &+2{\al'}'+2\al'f_\X+k_{\X t}+h_\X h_\Y-k_\X
h_\Z,\hspace{7.1cm}(4.58)
\end{eqnarray*}
\begin{eqnarray*}&&
[(f_\X-(-1)^r\gm\nu)\vf+\vf')\vt_r
+((f_\X-(-1)^r\nu\gm)\sgm+\sgm')\td\varpi\phi_r
-\al_1\gm\tau\zeta_r]_\Z
\\ &=&2(\be'\sin\al)'+2\be'f_\X\sin\al+h_{\X t}-k_\Y h_\X
+k_\X k_\Z \\
& &+2[(\gm_2\sgm)'-\gm_2\sgm(k_\Y+(-1)^r\nu\gm)+\gm_1\sgm k_\Z
]\phi_r. \hspace{6cm}(4.59)
\end{eqnarray*}
Thus we have:
$$2{\al'}'+2\al'f_\X+k_{\X t}+h_\X h_\Y-k_\X
h_\Z=0,\eqno(4.60)$$
$$2(\be'\sin\al)'+2\be'f_\X\sin\al+h_{\X t}-k_\Y h_\X
+k_\X k_\Z=0.\eqno(4.61)$$

For simplicity, we only consider two special cases a follows. \psp

{\it Case 1}. $\vt_r=\zeta_r,\;\sgm=0,\;\gm_1=\al'\mu$ and
$\gm_2=\be'\mu\sin\al$.\psp

In this case,
$$ k_\X=h_\X=0\eqno(4.62)$$
by (4.52). According to (4.60) and (4.61), we have
$$\be'\sin\al=d\al',\qquad
f_\X=-\frac{{\al'}'}{\al'}.\eqno(4.63)$$ Moreover, (4.51) becomes
$$k_\Y+dh_\Y=-\frac{(\mu\al')'}{\mu\al'},\qquad
k_\Z+dh_\Z=-d\frac{(\mu\al')'}{\mu\al'}.\eqno(4.64)$$ According to
(4.40) and (4.54),
$$h_\Z=\frac{{\al'}'}{\al'}-k_\Y,\qquad
h_\Y=k_\Z-2\be'\cos\al.\eqno(4.65)$$ Substituting (4.65) into
(4.64), we obtain
$$k_\Y+dk_\Z=2d\be'\cos\al-\frac{(\mu\al')'}{\mu\al'},\qquad
k_\Z-dk_\Y=-d\frac{(\mu\al')'+{\al'}'\mu}{\mu\al'}.\eqno(4.66)$$

For convenience of computation, we write
$$\mu=\frac{\sqrt{\be_1'}}{\al'}\lra\gm=(1+d^2)\be_1',\;\;\gm_1=
\sqrt{\be_1},\;\;\gm_2=d\sqrt{\be_1}.\eqno(4.67)$$ From (4.66),
$$k_\Y=\frac{1}{1+d^2}\left(2d^2\al'\cot\al+
(d^2-1)\frac{{\be_1'}'}{2\be_1'}+d^2\frac{{\al'}'}{\al'}\right),
\eqno(4.68)$$
$$k_\Z=\frac{d}{1+d^2}\left(2d^2\al'\cot\al-\frac{{\be_1'}'}{\be_1'}-\frac{{\al'}'}{\al'}\right).
\eqno(4.69)$$ By (4.65),
$$h_\Z=\frac{1}{1+d^2}\left((1-d^2)\frac{{\be_1'}'}{2\be_1'}
+\frac{{\al'}'}{\al'}-2d^2\al'\cot\al\right),\eqno(4.70)$$
$$h_\Y=-\frac{d}{1+d^2}\left(2\al'\cot\al
+\frac{{\be_1'}'}{\be_1'}+\frac{{\al'}'}{\al'}\right).
\eqno(4.71)$$ Furthermore, (4.57) becomes
$$(\gm\tau)'+2\gm\tau f_\X-(-1)^r\nu
\gm^2\tau=0\lra\gm\tau=(\al')^2e^{(-1)^r\nu(1+d^2)\be_1}.\eqno(4.72)$$
So $$ \tau=\frac{(\al')^2e^{(-1)^r\nu(1+d^2)\be_1}}
{(1+d^2)\be_1'}.\eqno(4.73)$$
 Note that (4.58) and (4.59) are implied by
$$(f_\X-(-1)^r\gm\nu)\vf+\vf'-\al_1\gm\tau=0\lra
\al_1=\frac{\al'\vf'-({\al'}'+(-1)^r\nu(1+d^2)\al'\be_1')\vf}
{(\al')^3e^{(-1)^r\nu(1+d^2)\be_1}}.\eqno(4.74)$$

Observe
\begin{eqnarray*}\hspace{2cm}\U&=&-\frac{{\al'}'}{\al'}\X-\al'(\Y+d\Z)
+\frac{\al'\vf'-({\al'}'+(-1)^r\nu(1+d^2)\al'\be_1')\vf}
{(\al')^3e^{(-1)^r\nu(1+d^2)\be_1}}\\
& &+(\vf-(\al')^2e^{(-1)^r\nu(1+d^2)\be_1}
\X)\zeta_r,\hspace{6.1cm}(4.75)\end{eqnarray*}
\begin{eqnarray*}\V&=&
\frac{1}{1+d^2}\left(2d^2\al'(\Y+d\Z)\cot\al+
((d^2-1)\Y-2d\Z)\frac{{\be_1'}'}{2\be_1'}+(d^2\Y-d\Z)\frac{{\al'}'}
{\al'}\right)\\
& &+\al'(\X-d\Z\cot\al)+\frac{(\al')^2e^{(-1)^r\nu(1+d^2)\be_1}}
{(1+d^2)\sqrt{\be_1'}}\xi_r,\hspace{5.8cm}(4.76)\end{eqnarray*}
\begin{eqnarray*}\W&=&\frac{1}{1+d^2}\left(((1-d^2)\Z-2d\Y)\frac{{\be_1'}'}{2\be_1'}
+(\Z-d\Y)\frac{{\al'}'}{\al'}-2d(\Y+d\Z)\al'\cot\al\right)
\\ & &+d\al'(\X+\Y\cot\al)+\frac{d(\al')^2e^{(-1)^r\nu(1+d^2)\be_1}}
{(1+d^2)\sqrt{\be_1'}}\xi_r.\hspace{5.5cm}(4.77)\end{eqnarray*}
Moreover,
\begin{eqnarray*}
\hspace{1cm}R_1&=&(4b\dlt_{0,r}+c\dlt_{1,r})c(\al')^2e^{(-1)^r\nu(1+d^2)\be_1}[
(\al')^2e^{(-1)^r\nu(1+d^2)\be_1}\X -\vf]\\
&&-\frac{(2({\al'}')^2-{{\al'}'}')\X}{(\al')^2}-\frac{2(\al')^3e^{(-1)^r\nu(1+d^2)\be_1}}
{\sqrt{\be_1'}}\xi_r-(\al')^2(1+d^2)\X\\ & &
+d((\al')^2\cot\al-{\al'}')\Z-({\al'}'+(d\al')^2\cot\al)\Y
,\hspace{4cm}(4.78)
\end{eqnarray*}
\begin{eqnarray*}R_2&=&\gm_1\left[\frac{\al'(2{\al'}'\be_1'-\al'{\be_1'}')e^{(-1)^r\nu(1+d^2)\be_1}}
{(1+d^2)(\be_1')^2}\xi_r+\frac{(\al')^4e^{(-1)^r2\nu(1+d^2)\be_1}}
{(1+d^2)\be_1'}\xi_r\zeta_r\right]\\ &&
-({\al'}'+(d\al')^2\cot\al)\X +[k_{\Y
t}-(\al')^2(1+d^2\csc^2\al)]\Y+\frac{1}{2}(k^2+h^2)_\Y
\\ & &+[(k_\Z-\be'\cos\al)'-d(\al')^2]\Z+2\al'f^\ast
-2(\al')^3e^{(-1)^r\nu(1+d^2)\be_1}\zeta_r\X,
\hspace{1.8cm}(4.79)\end{eqnarray*}
\begin{eqnarray*}R_3&=&\gm_2\left[\frac{\al'(2{\al'}'\be_1'-\al'{\be_1'}')e^{(-1)^r\nu(1+d^2)\be_1}}
{(1+d^2)(\be_1')^2}\xi_r+\frac{(\al')^4e^{(-1)^r2\nu(1+d^2)\be_1}}
{(1+d^2)\be_1'}\xi_r\zeta_r\right]\\ && +
d((\al')^2\csc\al-{\al'}')\X+((h_\Y+\be'\cos\al)'-d(\al')^2)\Y
+\frac{1}{2}(k^2+h^2)_\Z\\
& &+(h_{\Z t}-(d\al')^2\csc^2\al)\Z+2d\al'f^\ast
-2d(\al')^3e^{(-1)^r\nu(1+d^2)\be_1}\zeta_r\X.
\hspace{2.4cm}(4.80)
\end{eqnarray*}
By (3.12), \begin{eqnarray*}& &p
=\rho\{\frac{2(\al')^3e^{(-1)^r\nu(1+d^2)\be_1}\xi_r\X}
{\sqrt{\be_1'}}+\frac{(\al')^2}{2}(\Y^2+d^2\csc^2\al(\Y^2+\Z^2)+2d\Y\Z)
-\frac{2\al'\vf\zeta_r}{\sqrt{\be_1'}}\\
& &+\frac{(2({\al'}')^2-{{\al'}'}')\X^2} {2(\al')^2}
+\frac{\al'{{\al'}'}'-({\al'}')^2}{2(1+d^2)\al'}
(2d\Y\Z-d^2\Y^2-\Z^2)
-\frac{(\al')^4e^{(-1)^r2\nu(1+d^2)\be_1}\xi_r^2} {2(1+d^2)\be_1'}
\\ & &
+\frac{(-1)^r\al'(\al'{\be_1'}'-2{\al'}'\be_1') \zeta_r}
{(1+d^2)(\be_1')^2e^{(-1)^{r+1}\nu(1+d^2)\be_1}} +
\frac{2\al'[({\al'}'+(-1)^r\nu(1+d^2)\al'\be_1')\vf-\al'\vf'](\Y+d\Z)}
{(\al')^3e^{(-1)^r\nu(1+d^2)\be_1}}
\\& &-\frac{1}{2(1+d^2)^2}[\left(2d^2\al'(\Y+d\Z\cot\al)+
\frac{{\be_1'}'((d^2-1)\Y-2d\Z)}{2\be_1'}+\frac{{\al'}'(d^2\Y-d\Z)}
{\al'}\right)^2\\& &+
\left(\frac{{\be_1'}'((1-d^2)\Z-2d\Y)}{2\be_1'}
+\frac{{\al'}'(\Z-d\Y)}{\al'}-2d(\Y+d\Z)\al'\cot\al\right)^2]-\frac{(\al')^2(1+d^2)\X^2}{2}
\\ &&+\frac{d}{1+d^2}({\al'}'\cot\al-(\al')^2\csc^2\al)((1-d^2)\Y\Z+2d(\Z^2-
\Y^2))
+\frac{\be_1'{{\be_1'}'}'-({\be_1'}')^2} {4(1+d^2)(\be_1')^2}\\
& &\times(4d\Y\Z+(1-d^2)(\Y^2-\Z^2))
+({\al'}'+(d\al')^2\cot\al)\X\Y +d({\al'}'-(\al')^2\cot\al)\X\Z
\hspace{4cm}\end{eqnarray*}
\begin{eqnarray*}&
&+(4b\dlt_{0,r}+c\dlt_{1,r})c(\al')^2e^{(-1)^r\nu(1+d^2)\be_1}[\vf\X-
2^{-1}(\al')^2e^{(-1)^r\nu(1+d^2)\be_1}\X^2 ]\} \hspace{2cm}(4.81)
\end{eqnarray*}
modulo the transformation in (1.10).

By (3.3) and (3.5), we have the following theorem: \psp

{\bf Theorem 4.2}. {\it Let $\al,\be_1,\vf$ be functions of $t$
and let $b,c,d$ be real constants. Denote
$$\be=d\ln|\csc\al-\cot\al|\eqno(4.82)$$
(so the first equation in (4.63) holds). Define the moving frame
$\X,\;\Y$ and $\Z$ by (3.1) and (3.5), and
$$\xi_0=be^{\sqrt{\be_1}(\Y+d\Z)}-ce^{-\sqrt{\be_1}(\Y+d\Z)},\qquad
\xi_1=c\sin[\sqrt{\be_1}(\Y+d\Z)+b],\eqno(4.83)$$
$$\zeta_0=be^{\sqrt{\be_1}(\Y+d\Z)}+ce^{-\sqrt{\be_1}(\Y+d\Z)},\qquad
\zeta_1=c\cos[\sqrt{\be_1}(\Y+d\Z)+b].\eqno(4.84)$$ For $r=0,1$,
we have the following solution of the Navier-Stokes equations
(1.1)-(1.4):
\begin{eqnarray*}u&=&\left[\frac{\al'\vf'-({\al'}'+(-1)^r\nu(1+d^2)\al'\be_1')\vf}
{(\al')^3e^{(-1)^r\nu(1+d^2)\be_1}}+(\vf-(\al')^2e^{(-1)^r\nu(1+d^2)\be_1}\X)
\zeta_r-\frac{{\al'}'}{\al'}\X\right]\cos\al
\\& &-\frac{1}{1+d^2}\left(2d^2\al'(\Y+d\Z)\cot\al+
((d^2-1)\Y-2d\Z)\frac{{\be_1'}'}{2\be_1'}+(d^2\Y-d\Z)\frac{{\al'}'}
{\al'}\right)\sin\al\\
& &+\al'(\X\sin\al-(\Y+2d\Z)\cos\al)
+\frac{(\al')^2e^{(-1)^r\nu(1+d^2)\be_1}}
{(1+d^2)\sqrt{\be_1'}}\xi_r\sin\al,\hspace{2.7cm}(4.85)\end{eqnarray*}
\begin{eqnarray*}v&=&\left[\frac{\al'\vf'-({\al'}'+(-1)^r\nu(1+d^2)\al'\be_1')\vf}
{(\al')^3e^{(-1)^r\nu(1+d^2)\be_1}}+(\vf-(\al')^2e^{(-1)^r\nu(1+d^2)\be_1}\X)
\zeta_r\right]\sin\al\;\cos\be
\\& &+\frac{2d\al'(d\cos\al\;\cos\be-\sin\be)(\Y+d\Z)\cot\al}
{1+d^2}+\frac{{\al'}'(d\cos\al\;\cos\be+\sin\be)(d\Y-\Z)}
{(1+d^2)\al'}\\ & &
+\frac{[((d^2-1)\Y-2d\Z)\cos\al\;\cos\be-((1-d^2)\Z-2d\Y)\sin\be]{\be_1'}'}
{2(1+d^2)\be_1'}-\frac{{\al'}'}{\al'}\X\sin\al\;\cos\be\\ &
&+\al'\X(\cos\al\;\cos\be-d\sin\be)-\al'\Y(\cos\al\;\cos\be-d\cos\al\;\sin\be)\\
& &-d\al'\Z\csc\al\;\cos\be
+\frac{(\al')^2e^{(-1)^r\nu(1+d^2)\be_1}(\cos\al\;\cos\be-d\sin\be)}
{(1+d^2)\sqrt{\be_1'}}\xi_r,\hspace{2.1cm}(4.86)
\end{eqnarray*}
\begin{eqnarray*}w&=&\left[\frac{\al'\vf'-({\al'}'+(-1)^r\nu(1+d^2)\al'\be_1')\vf}
{(\al')^3e^{(-1)^r\nu(1+d^2)\be_1}}+(\vf-(\al')^2e^{(-1)^r\nu(1+d^2)\be_1}\X)
\zeta_r\right]\sin\al\;\sin\be
\\& &+\frac{2d\al'(d\cos\al\;\sin\be+\cos\be)(\Y+d\Z)\cot\al}
{1+d^2}+\frac{{\al'}'(d\cos\al\;\sin\be-\cos\be)(d\Y-\Z)}
{(1+d^2)\al'}\\ & &
+\frac{[((d^2-1)\Y-2d\Z)\cos\al\;\sin\be+((1-d^2)\Z-2d\Y)\cos\be]{\be_1'}'}
{2(1+d^2)\be_1'}-\frac{{\al'}'}{\al'}\X\sin\al\;\sin\be\\ &
&+\al'\X(\cos\al\;\sin\be+d\cos\be)-\al'\Y(\cos\al\;\sin\be+d\cos\al\;\cos\be)\\
& &-d\al'\Z\csc\al\;\sin\be
+\frac{(\al')^2e^{(-1)^r\nu(1+d^2)\be_1}(\cos\al\;\sin\be+d\cos\be)}
{(1+d^2)\sqrt{\be_1'}}\xi_r\hspace{2.2cm}(4.87)
\end{eqnarray*}
and $p$ is given in (4.81).}\psp

{\it Case 2}. $\gm_2=\al_1=0,\;(\gm\tau)'+2\gm\tau f_\X-(-1)^r\nu
\gm^2\tau=0$ and $\gm_1\neq 0$.\psp

According to (4.51) and (4.54),
$$k_\Y=-\frac{\gm_1'}{\gm_1},\qquad k_\Z=0,\qquad
h_\Y=-2\be'\cos\al.\eqno(4.88)$$ Note $\gm=\gm_1^2$. Moreover,
(4.52) says
$$k_\X=0,\qquad h_\X=-2\be'\sin\al.\eqno(4.89)$$
Furthermore, (4.61) yields
$$f_\X=\frac{\gm_1'}{\gm_1}.\eqno(4.90)$$
Besides, (4.40) implies
$$h_\Z=-(f_\X+k_\Y)=0.\eqno(4.91)$$
Under the condition (4.60) and (4.61), (4.58) and (4.59) are
equivalent to
$$f_\X=k_\Y=\gm_1'=0,\qquad -(-1)^r\gm\nu\vf+\vf'=0.\eqno(4.92)$$

 Write $\gm_1=a_1$ as a real constant. We have:
$$\sgm=a_2e^{(-1)^r\nu a_1^2t},\qquad
\vf=a_1e^{(-1)^r\nu a_1^2t},\qquad\tau =a_1^{-1}e^{(-1)^r\nu
a_1^2t} \eqno(4.93)$$ for $a_2\in\mbb{R}$ (cf. (4.47)). By (4.60)
$$\be'=\pm
\sqrt{\frac{-{\al'}'}{\sin2\al}},\eqno(4.94)$$ that is,
$$\be=\pm\int
\sqrt{\frac{-{\al'}'}{\sin2\al}}\;dt.\eqno(4.95)$$
 Thus
$$\U=-\al'\Y-\be'\Z\sin\al+a_1e^{(-1)^r\nu
a_1^2t}(\vt_r+a_2\Y\phi_r-\zeta_r\X-a_1a_2\psi_r\X^2),\eqno(4.96)$$
$$\V=\al'\X-\be'\Z\cos\al+e^{(-1)^r\nu
a_1^2t}(\xi_r+2a_1a_2\phi_r\X),\qquad\W=-\be'(\X\sin\al+\Y\cos\al).
\eqno(4.97)$$ Moreover,
\begin{eqnarray*}&&
R_1=-\frac{{\al'}'}{2}\Y+a_1^2e^{(-1)^r2\nu
a^2_1t}[(4b\dlt_{0,r}+c\dlt_{1,r})c\X+3a_2(2\dlt_{0,r}(ab+c)+
\dlt_{1,r}c\cos b)\X^2\\ & &+2a_1^2a_2^2
(4a\dlt_{0,r}+\dlt_{1,r})\X^3
+a_2(2\dlt_{0,r}(ab-c)+\dlt_{1,r}c\sin b)\Y+
\dlt_{1,r}cc_1\sin(b-b_1)\\
& &+2\dlt_{0,r}(bc_1-b_1c)+a_1a_2(2\dlt_{0,r}(c_1-ab_1)-
\dlt_{1,r}c_1\sin b_1)\X] +(\al'\be'\cos\al-(\be'\sin\al)')\Z
\\ & &-((\al')^2+(\be'\sin\al)^2)\X + e^{(-1)^r\nu
a^2_1t}(a_1a_2\phi-2\al') ) (\xi_r+2a_1a_2e^{(-1)^r\nu
a^2_1t}\phi_r\X),\hspace{0.8cm}(4.98)
\end{eqnarray*}
\begin{eqnarray*}R_2&=&2a_1e^{(-1)^r\nu
a_1^2t}[(\vt_r+a_2\Y\phi_r)(\al'+a_1a_2e^{(-1)^r\nu
a_1^2t}\phi_r)-\al'(\zeta_r\X+a_1a_2\psi_r\X^2)]\\ & &
-\frac{{\al'}'}{2}\X
 +(3(\be')^2\cos^2\al-(\al')^2)\Y
-((\be'\cos\al)'+\al'\be'\sin\al)\Z\\&& +a_1e^{(-1)^r2\nu a_1^2t}(
\xi_r\zeta_r+2a_1a_2\xi_r\psi_r\X+2a_1^2a_2^2\phi_r\psi_r\X^2)
,\hspace{4.2cm}(4.99)\end{eqnarray*}
$$R_3=(\al'\be'\cos\al-(\be'\sin\al)')\X
-((\be'\cos\al)'+\al'\be'\sin\al)\Y+{\al'}'\Z\csc2\al.\eqno(4.100)$$
By (3.12),
\begin{eqnarray*}&&p
=\rho\{\frac{{\al'}'\X\Y}{2}-a_1^2e^{(-1)^r2\nu
a^2_1t}[\frac{(4b\dlt_{0,r}+c\dlt_{1,r})c\X^2}{2}
+a_2(2\dlt_{0,r}(ab+c)+ \dlt_{1,r}c\cos b)\X^3\\ & &+a_1^2a_2^2
(4a\dlt_{0,r}+\dlt_{1,r})\frac{\X^4}{2}
+a_2(2\dlt_{0,r}(ab-c)+\dlt_{1,r}c\sin b)\X\Y+
(\dlt_{1,r}cc_1\sin(b-b_1)\\
& &+2\dlt_{0,r}(bc_1-b_1c))\X+\frac{a_1a_2(2\dlt_{0,r}(c_1-ab_1)-
\dlt_{1,r}c_1\sin b_1)\X^2}{2}]
+\frac{((\al')^2+(\be'\sin\al)^2)\X^2}{2}\\
& &+((\be'\sin\al)'-\al'\be'\cos\al)\X\Z- e^{(-1)^r\nu
a^2_1t}(a_1a_2\phi-2\al')(\xi_r\X+a_1a_2e^{(-1)^r\nu
a^2_1t}\phi_r\X^2)\\& & - (-1)^re^{(-1)^r\nu a_1^2t}[
a_2e^{(-1)^r\nu
a_1^2t}[a_1a_2\Y\phi_r\psi_r-2^{-1}a_2(a_1(4a\dlt_{0,1}+\dlt_{1,r})\Y^2+
\phi_r^2)\\ &
&+\vt_r\zeta_r-(\dlt_{0,1}2(ab_1+c_1)+\dlt_{1,r}c_1)\Y
]+\al'(\ves_r+a_2\Y\psi_r-a_1^{-1}a_2\phi_r)]-\frac{e^{(-1)^r2\nu
a_1^2t} \xi_r^2}{2}\\ & &
-\frac{(3(\be')^2\cos^2\al-(\al')^2)\Y^2}{2}
+((\be'\cos\al)'+\al'\be'\sin\al)\Y\Z
-\frac{{\al'}'\Z^2\csc2\al}{2}
\}\hspace{0.9cm}(4.101)\end{eqnarray*} where modulo the
transformation in (1.10), where
$$\ves_0=b_1e^{\td\varpi}+c_1e^{-\td\varpi},\qquad
\ves_1=c_1\cos(\td\varpi+b_1)\eqno(4.102)$$ in connection with
(4.56) and $\td\varpi=a_1\Y$.

By (3.3) and (3.5), we have the following theorem: \psp

{\bf Theorem 4.3}. {\it Let $\al$ be a function of $t$ and let
$a,a_1,a_2,b,b_1,c,c_1$ be real constants. Denote $\be$ as in
(4.91). Define the moving frame $\X,\;\Y$ and $\Z$ by (3.1) and
(3.5), and
$$\phi_0=
e^{a_1\Y}-ae^{-a_1\Y},\;\;\phi_1=\sin(a_1\Y),\;\;\psi_0=
e^{a_1\Y}+ae^{-a_1\Y},\;\; \psi_1=\cos(a_1\Y),\eqno(4.103)$$
$$\xi_0=
be^{a_1\Y}-ce^{-a_1\Y},\;\;\xi_1=c\sin(a_1\Y+b),\;\;\zeta_0=
be^{a_1\Y}+ce^{-a_1\Y},\eqno(4.104)$$ $$\zeta_1=c\cos(a_1\Y+b),
\;\;\vt_0=
b_1e^{a_1\Y}-c_1e^{-a_1\Y},\;\;\vt_1=c_1\sin(a_1\Y+b_1),\eqno(4.105)$$
$$\ves_0=
b_1e^{a_1\Y}+c_1e^{-a_1\Y},\qquad\ves_1=c_1\cos(a_1\Y+b_1).\eqno(4.106)$$
For $r=0,1$, we have the following solution of  Navier-Stokes
equations (1.1)-(1.4):
\begin{eqnarray*}\hspace{1cm}u&=&
[-\al'\Y+a_1e^{(-1)^r\nu
a_1^2t}(\vt_r+a_2\Y\phi_r-\zeta_r\X-a_1a_2\psi_r\X^2)]\cos\al
\\ & &-[\al'\X+e^{(-1)^r\nu
a_1^2t}(\xi_r+2a_1a_2\phi_r\X)]\sin\al,\hspace{5cm}(4.107)\end{eqnarray*}
\begin{eqnarray*}& &v=
[-\al'\Y+a_1e^{(-1)^r\nu
a_1^2t}(\vt_r+a_2\Y\phi_r-\zeta_r\X-a_1a_2\psi_r\X^2)]\sin\al\;\cos\be
-\be'\Z\cos\be\\ & &-[\al'\X+e^{(-1)^r\nu
a_1^2t}(\xi_r+2a_1a_2\phi_r\X)]\cos\al\;\cos\be
+\be'(\X\sin\al+\Y\cos\al)\sin\be
,\hspace{0.4cm}(4.108)\end{eqnarray*}
\begin{eqnarray*}& &w=
[-\al'\Y+a_1e^{(-1)^r\nu
a_1^2t}(\vt_r+a_2\Y\phi_r-\zeta_r\X-a_1a_2\psi_r\X^2)]\sin\al\;\sin\be
-\be'\Z\sin\be\\ & &-[\al'\X+e^{(-1)^r\nu
a_1^2t}(\xi_r+2a_1a_2\phi_r\X)]\cos\al\;\sin\be
-\be'(\X\sin\al+\Y\cos\al)\cos\be
\hspace{0.6cm}(4.109)\end{eqnarray*} and $p$ is given in (4.101).}

 \vspace{0.7cm}

\noindent{\Large \bf References}

\begin{description}

\item[{[BK]}] M. A. Brutyan and P. L. Krapivsky, Exact solutions
of Navier-Stokes equations describing the evolution of a vortex
structure in generalized shear flow, {\it Comput. Math. Phys.}
{\bf 32} (1992), 270-272.

\item[{[Ba]}] A. A. Buchnev, Lie group admitted by the equations
of motion of an ideal incompressible fluid, {\it Dinamika
Sploshnoi Sredi. Int. of Hydrodynamics Novosibirsk} {\bf 7}
(1971), 212.

\item[{[Bv1]}] V. O. Bytev, Nonsteady motion of a rotating ring of
viscous incompressible fluid with free boundary, {\it Zhumal
Prikladnoi Mekhaniki i Tekhnicheskoi Fiziki} {\bf 3} (1970), 83.

\item[{[Bv2]}] V. O. Bytev, Invariant solutions of the
Navier-Stokes equations, {\it Zhumal Prikladnoi Mekhaniki i
Tekhnicheskoi Fiziki} {\bf 6} (1972), 56.

\item[{[G]}] V. I. Gryn, Exact solutions of Navier-Stokes
equations, {\it J. Appl. Math. Mech.} {\bf 55} (1991), 301-309.

\item[{[I]}] N. H. Ibragimov, {\it Lie Group Analysis of
Differential Equations}, Volume 2, CRC Handbook, CRC Press, 1995.

\item[{[J]}] G. B. Jeffery, {\it Philosophical Magazine, Ser.}
{\bf 6} (1915), 29.

\item[{[K]}] L. B. Kapitanskii, Group analysis of Navier-Stokes
equations and Euler equations with rotational symmetry and new
exact solutions of these equations, {\it Dokl. Akad. Nauk
S.S.S.R.} {\bf 243} (1978), 901.

\item[{[KKR]}] H. E. Kochin, I. A. Kibel' and N. V. Roze, {\it
Theoretical Hydromechanics,} Fizmatgiz, Moscow, 1963.

\item[{[Ll]}] L. Landau, A new exact solutions of  Navier-Stokes
equations, {\it C. R. (Doklady) Acad. Sci. URSS (N. S.)} {\bf 43}
(1944), 286-288.

\item[{[Lr]}] R. B. Leipnik, Exact solutions of  Navier-Stokes
equations by recursive series of diffusive equations, {\it C. R.
Math. Rep. Acad. Sci. Canada} {\bf 18} (1996), 211-216.

\item[{[LRT]}] C. C. Lin, E. Reissner and H. S. Tsien, On
two-dimensional non-steady motion of a slender body in a
compressible fluid, {\it J. Math. Phys.} {\bf 27} (1948), no. 3,
220.

\item[{[O]}] L. V. Ovsiannikov, {\it Group Analysis of
Differential Equations}, Academic Press, New York, 1982.

\item[{[Pa]}] A. D. Polyanin, Exact solutions of the Navier-Stokes
equations with generalized separation of variables, {\it Dokl.
Phys.} {\bf 46} (2001), 726-731.

\item[{[Pv1]}] V. V. Pukhnachev, Group properties of Navier-Stokes
equations in two-dimensional case, {\it Zhumal Prikladnoi
Mekhaniki i Tekhnicheskoi Fiziki} {\bf 1} (1960), 83.

\item[{[Pv2]}] V. V. Pukhnachev, Invariant solutions of
Navier-Stokes equations describing motions with free boundary,
{\it Dokl. Akad. Nauk S.S.S.R.} {\bf 202} (1972), 302.

\item[{[S1]}] H.-C. Shen, The theory of functions of a complex
variable under Dirac-Pauli representation and its application in
fluid dynamics I, {\it Appl. Math. Mech. (English Ed.)} {\bf 7}
(1986), 391-411.

\item[{[S2]}] H.-C. Shen, Exact solutions of Navier-Stokes
equations---the theory of functions of a complex variable under
Dirac-Pauli representation and its application in fluid dynamics
II, {\it Appl. Math. Mech. (English Ed.)} {\bf 7} (1986), 557-562.

\item[{[V]}] V. G. Vyskrebtsov, New exact solutions of
Navier-Stokes equations for axisymmetric self-similar fluid flows,
{\it J. Math. Sci. (New York)} {\bf 104} (2001), 1456-1463.

\item[{[W]}] Z. Warsi, {\it Fluid Dynamics}, CRC Press LTC, 1999.

\item[{[X1]}] X. Xu,  Stable-Range approach to the equation of
nonstationary transonic gas flows, {\it Quart. Appl. Math}, to
appear.

\item[{[X2]}] X. Xu, Parameter-Function approach to classical
 non-steady boundary Layer problems, {\it Preprint.}

\item[{[Y]}] A. Yu. Yakimov, Exact solutions of Navier-Stokes
equations in the presence of a vortex singularity on a ray, {\it
Dokl. Acad. Nauk SSSR} {\bf 276} (1984), 79-82.

\end{description}

\end{document}